\documentclass[%
 aip,
 amsmath,amssymb,
 reprint,%
]{revtex4-1}
\usepackage{graphicx}
\usepackage{dcolumn}
\usepackage{bm}

\usepackage[utf8]{inputenc}
\usepackage[T1]{fontenc}
\usepackage{mathptmx}
\usepackage{verbatim}
\usepackage{xcolor}

\begin{document}

\preprint{AIP/123-QED}

\title[A photogrammetric method for target monitoring inside the MEG II detector]{ A photogrammetric method for target monitoring inside the MEG II detector}

\author{G. Cavoto}
\altaffiliation[Also at ]{INFN Roma, P.le Aldo Moro 2, 00185 Rome, Italy.}
\affiliation{Physics Department Sapienza Universit\`a di Roma, P.le Aldo Moro 2, 00185 Rome, Italy.}
\author{G. Chiarello}%
\affiliation{INFN Roma, P.le Aldo Moro 2, 00185 Rome, Italy.}
\author{M. Hildebrandt}%
\affiliation{Paul Scherrer Institut, Forschungsstrasse 111, 5232 Villigen, Switzerland.}
\author{A. Hofer}%
\affiliation{Paul Scherrer Institut, Forschungsstrasse 111, 5232 Villigen, Switzerland.}
\author{K. Ieki}%
\affiliation{ICEPP, The University of Tokyo, 7-3-1 Hongo, Bunkyo-ku, Tokyo 113-0033, Japan}
\author{M. Meucci}%
\affiliation{Physics Department, Sapienza Universit\`a di Roma, P.le Aldo Moro 2, 00185 Rome, Italy.}
\author{S. Milana}%
\affiliation{INFN Roma, P.le Aldo Moro 2, 00185 Rome, Italy.}
\author{V. Pettinacci}%
\affiliation{INFN Roma, P.le Aldo Moro 2, 00185 Rome, Italy.}
\author{F. Renga}%
\email{francesco.renga@roma1.infn.it.}
\affiliation{INFN Roma, P.le Aldo Moro 2, 00185 Rome, Italy.}
\author{C. Voena}%
\affiliation{INFN Roma, P.le Aldo Moro 2, 00185 Rome, Italy.}

\date{\today}

\begin{abstract}
An automatic target monitoring method based on photographs taken by a CMOS photo-camera has been developed for the MEG II detector. The technique could be adapted for other fixed-target experiments requiring good knowledge of their target position to avoid biases and systematic errors in measuring the trajectories of the outcoming particles. A CMOS-based, high resolution, high radiation tolerant and high magnetic field resistant photo-camera was mounted inside the MEG II detector at the Paul Scherrer Institute (Switzerland). MEG II is used to search for lepton flavour violation in muon decays. The photogrammetric method's challenges, affecting measurements of low momentum particles' tracks, are high magnetic field of the spectrometer, high radiation levels, tight space constraints, and the need to limit the material budget in the tracking volume. The camera is focused on dot pattern drawn on the thin MEG II target, about 1 m away from the detector endcaps where the photo-camera is placed. Target movements and deformations are monitored by comparing images of the dots taken at various times during the measurement. The images are acquired with a Raspberry board and analyzed using a custom software. Global alignment to the spectrometer is guaranteed by corner cubes placed on the target support. As a result, the target monitoring fulfils the needs of the experiment.
\end{abstract}

\maketitle

\section{\label{sec:intro}Introduction}

Magnetic spectrometers used to determine the momentum of charged particles  in high-energy physics experiments (HEP) require an accurate reconstruction of the trajectory of the particle over a relatively large volume. This is usually achieved by measuring with high precision various  positions in space and then connecting them to obtain the best evaluation of the trajectory. The relative uncertainty on the momentum of a charged particle is equal to the relative uncertainty on the curvature of the trajectory.
Typical particle detectors used in HEP spectrometers are gaseous drift chambers, time-projection chambers, etc., which can be made of sub-elements that require an accurate relative alignment. Moreover, it is important to measure their relative position with respect to other elements, including a production target where the charged particles under study are emerging from.
Generally, to reach the desired performances, high accuracy of the mechanical assembly is required. However, due to the apparatuses' complexity, these measurements are often difficult and ad-hoc solutions need to be developed to reach sub-millimeter alignment of the critical elements. Space constraints, strong magnetic fields, and high radiation levels add to the list of challenges. Different solutions were adopted in HEP experiments, some of them exploiting optical detection of patterns printed on the detectors themselves~\cite{Beker_2019}. Similar challenges arise in high-power laser experiments, where targets have to be replaced in a fast turnaround time~\cite{laser}.

In this paper we describe a method that has been developed to monitor the position of the muon stopping target in the MEG II experiment at the Paul Scherrer Institute (PSI, Villigen, Switzerland). The method is based on a photogrammetric survey of a dot pattern printed on the target itself. The main challenges of this approach are connected to the use of photo cameras in an environment with high magnetic field and high radiation. Since these conditions are common in HEP experiments, the approach may be of interest to other experiments.

The MEG II experiment \cite{Baldini:2018nnn} is an upgrade of the MEG experiment, which set the best world limit\cite{TheMEG:2016wtm} on the decay of a muon into a positron and a photon, $\mu^+ \rightarrow e^+ \gamma$. This decay was searched for since the discovery of the muon, and never observed. Indeed, it is practically forbidden in the Standard Model of particle physics, and its discovery would be the demonstration of new physics effects. If the decay is not observed, MEG II is expected to set an upper limit of $6 \times 10^{-14}$ on its branching ratio, further constraining theoretical models for physics beyond the Standard Model. The future availability of higher intensity muon beams could further improve the experimental sensitivity to this decay\cite{Cavoto:2017kub}.

\begin{figure}[htbp]
\begin{center}
\includegraphics[width=0.55\textwidth, angle=0]{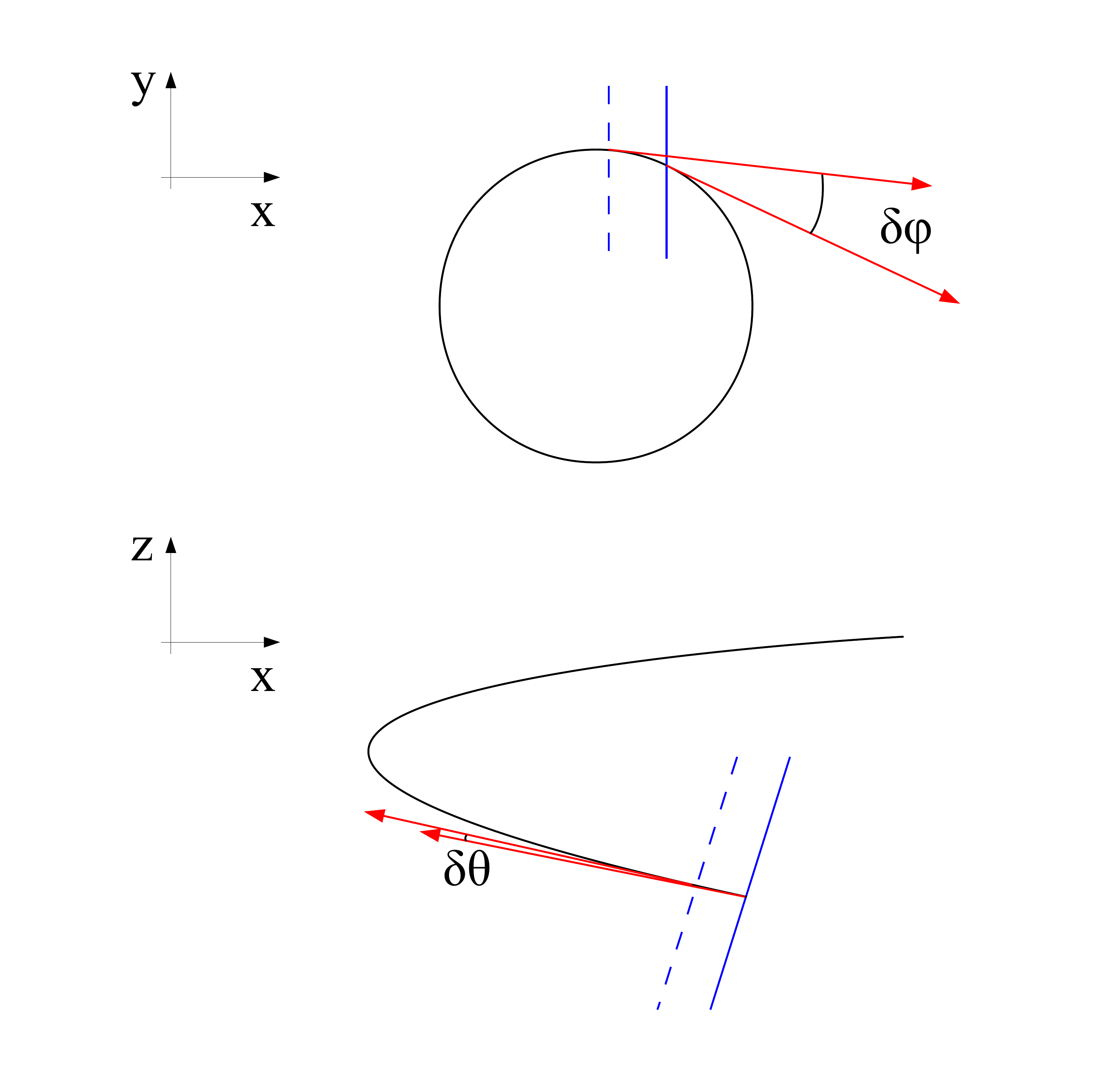}
\caption{Sketch (not in scale) of the impact of a target displacement on the reconstructed track angles. The dashed (full) segment represents the assumed (true) target projection in the corresponding plane. Top: projection on the x-y plane, where the positron trajectory projection is a circle. $\delta \phi$ is the difference between the true and the reconstructed azimuthal angle of the track. Bottom: projection on the x-z plane, the curve represents the positron trajectory projection in this plane. $\delta \theta$ is the difference between the true and the reconstructed polar angle of the track. }
\label{fig:thetaphi}
\end{center}
\end{figure}

The search for $\mu^+ \rightarrow e^+ \gamma$ requires stopping a large amount of muons, detecting a positron and a photon emerging in coincidence from the stopping target, and reconstructing their kinematics. In the MEG II experiment, the PSI beam of positive muons ($7 \times 10^7$ muons per second) is stopped in a thin plastic target at the center of the MEG II detector. It includes a spectrometer to measure the trajectory of the 52.8 $\rm{MeV}$ positrons possibly produced in the $\mu^+ \rightarrow e^+ \gamma$ decay, and a liquid Xenon (LXe) calorimeter to detect the photon (plus some auxiliary detectors). The MEG II magnetic spectrometer is composed of a single volume multi-wire drift chamber \cite{Chiarello:2019cvx,Baldini:2020rds} in a solenoidal gradient magnetic field. One of the dominant systematic errors in the evaluation of the yield of $\mu^+ \rightarrow e^+ \gamma$ events  in the previous MEG experiment was due to the uncertainty on the target position with respect to the spectrometer and its internal deformation which could not be measured directly.
In order to identify a  $\mu^+ \rightarrow e^+ \gamma$ event, it is necessary to measure the angles of the $e^+$  trajectory at the point where the muon has decayed (muon decay point).  This is done by back-propagating  the trajectory measured by the spectrometer up to the target region, that is assumed to be a planar surface. The MEG II spectrometer is expected to provide a precision of about 5~mrad on the $\theta$ (polar) and $\phi$ (azimuthal) angles of the positron trajectory at the target. In the MEG II   reference system  $z$ is the axis along the beam direction. A precise knowledge of the target position is then required: given a radius of curvature of about 13 cm for the $e^+$ trajectory in  $\mu^+ \rightarrow e^+ \gamma$ events, a displacement of the target by 500~$\mu$m along the  direction  normal to it implies a systematic deviation of about 4~mrad in the measured positron $\phi$ angle for $\phi = 0$. A even larger effect is expected for non-zero values of $\phi$. Figure~\ref{fig:thetaphi} shows the effect of a displacement of the target in the direction orthogonal to its plane, on the reconstructed track angles. It can be seen that the impact on the azimuthal angle can be sizable while the effect on the polar angle is negligible.
Moreover, deformations of the target planarity, which produce a similar effect, were observed through the MEG data taking. The uncertainty on the target position and deformation was in fact the dominant systematic error on the MEG result. It caused a $5\%$ variation of the upper limit on the branching fraction while other contributions were below $1\%$.

During the MEG data-taking the position of the target plane was measured every run period (i.e. every year) with an optical survey of crosses depicted on the target plane. Unfortunately the small field of view available for triangulation, combined with the distance of the target from the closest accessible point of view (about 1 m), prevented to achieve an accuracy better than 1 mm. Also, a target position monitoring over long data taking periods was possible by reconstructing the position of a few holes bored on the
target itself. A map of the reconstructed muon decay points on the target clearly showed the position of such holes. If the target position assumed
in the trajectory reconstruction procedure is not exact, the holes artificially appear at different positions for different $e^+$ angles. This allowed
to reconstruct deviations of the target position from the nominal one. This method was also effective  to catch and correct
the deformation of the target planarity. On the other hand, it required a large amount of data, so that it could only be
used to monitor the average target position over a few months of data taking. However,  the target was removed far from its working position at least every week
to perform the calibration of the LXe detector.
 A pneumatic
system was used for this, but it did not ensure a micrometric repeatability of the target positioning. While the target hole technique 
was precise enough for the MEG experiment, the improved resolutions of the MEG II positron spectrometer imposed the development of an 
alternative method. It must ensure a  more frequent monitoring of the target position over the data taking period and  has to be able to resolve
displacements equivalent to about 100~$\mu$m along the  direction  normal to the target plane.

We here present a photogrammetric approach which will employ a digital
CMOS photo-camera to take pictures of a pattern drawn on the target itself. The photo-camera will be placed in the inner cavity of the MEG II cylindrical drift chamber where muons travel along   to reach the target. The engineering of the photo-camera mounting  will play a key role: it must ensure dimensional mechanical stability over time in a high radiation environment  and sufficient rigidity to adequately support the instrumentation. The support - although necessarily compact in size -  should avoid deformation and should not be affected by the high active magnetic field. All this requires a study of   non-magnetic materials to be used.  Moreover, the total amount of material should be kept as small as possible, because positrons hitting the system can produce photon background in the calorimeter. We will show that the photo-camera can be installed without affecting the muon propagation and the magnetic field. Together with the photo-camera described in this paper, a different photo-camera was installed and tested inside the MEG II detector~\cite{Bill}. Here, we propose different optical configuration and algorithms. Moreover, a systematic analysis of the achievable resolution, obtained in a controlled bench-top set-up, will be presented.

\section{\label{sec:approach}The photogrammetric approach}
\subsection{\label{subsec:setup}The experimental setup}

\begin{figure}[htbp]
\begin{center}
\includegraphics[width=0.45\textwidth]{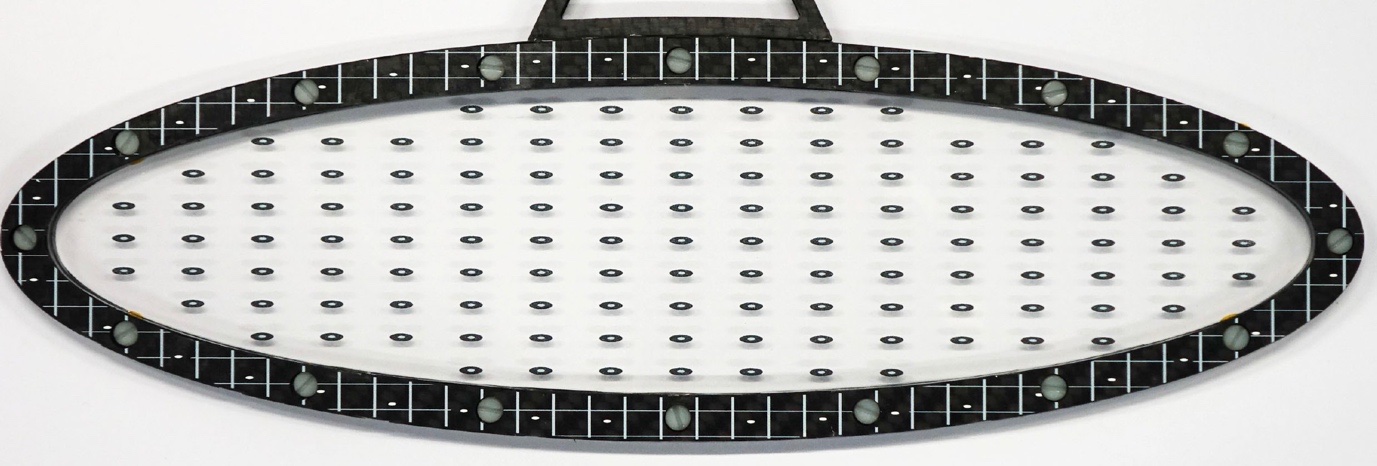}
\caption{The MEG II target with the dot pattern on the foil and the frame.}
\label{fig:pattern}
\end{center}
\end{figure}

The MEG II target is an elliptical foil (length of 270 mm and height of 66 mm) with 174~$\mu$m average thickness, made of scintillating material. Its normal direction lies on the horizontal plane and forms an angle of $75^\circ$ with respect to the $z$ axis.
The target foil is supported by two hollow carbon fiber frames. 
A pattern of white dots, superimposed on a black background, is printed on both the frame and the foil. 
The dots are elliptical with an height and a width of 0.51 mm and 1.52 mm on the target and 0.42 mm and 
1.27 mm on the frame. The ratio of the two axes is chosen in such a way that, considering
the target orientation and the photo-camera position, the dots look circular in the picture. This dots pattern has been found superior to others including white lines with 
black contours. Figure~\ref{fig:pattern} shows a picture of the MEG II target.
%
The dots are imaged with a digital CMOS photo-camera (IDS, mod. UI-3282SE), with a Sony IMX264 sensor having 
$2456 \times 2054$ pixels of 3.5~$\mu$m size, for a total sensor size of 8.473~mm $\times$ 7.086~mm. 
A TUSS optical system, mod. LVK7518, with a focal length of 75~mm and a maximum aperture of $f/1.8$ is used. The read-out of the photo-camera uses a Raspberry board, hosted in a crate close to the apparatus, that is using a USB3 protocol for the communication. Previous tests found that the Ethernet communication is not compatible with the MEG II high magnetic field. The USB cable from the photo-camera, which also provides power, exits the internal volume via a feed-through present on a connection flange, sealed with glue. A LED system supported independently illuminates the target
during the data-taking.

The value of the  magnetic field at the  position where the photo-camera is installed  along the $z$ axis is about 0.8 T.  The magnetic force applied on the photo-camera assembly has been evaluated and the support design  optimized through a dedicated topological analysis in terms of material and geometry.  This  ensures a proper rigidity during the measurement stage.
The possible interference  induced by  the photo-camera to the magnetic field was measured with  Hall probes and found to be negligible.
The photo-camera was placed on an ad-hoc support, approximately in the nominal position, which hosted a Hall cube with 6 sensors (2 for each direction).
The observed deviations from the total field in the presence of the photo-camera were found to be less than 0.6$\%$, 0.2$\%$, 0.2$\%$ in the $x$, $y$, $z$ directions respectively.

To evaluate the effect of radiation damage, a photo-camera with the same sensor was left installed  for more than one month during the 2017  MEG II engineering run. Although  an increase in the number of hot pixels was observed, the effect is far too low to 
affect significantly the performances of the measurement system. 

In conclusion, we are confident that the final photo-camera will work inside 
the COBRA magnetic field, will not affect the field itself and will remain operative for the expected time of MEG II data-taking (3 years).

The  photogrammetric approach is based on the repetition of several  measurements  of the same points at different times. Therefore the stability of the photo-camera mounting is crucial. Furthermore, the correct positioning is fundamental even to avoid any interference with the muon beam entering the multi-wire chamber and other equipment installed in that area. The space allocated for the instrumentation is outside the tracking volume, in order not to interfere with the positron detection.  A clearance of 80~mm around the beam axis was left  in order not to intersect the beam halo.
Within the allowed space, it was necessary to define a system capable of aiming with extreme precision at the centre of the target. Given the small space available, it was not possible to insert pointing adjustment elements and a solution with a fixed setup was chosen. Considering also the need of inserting as little material as possible and the complexity of the  shape, it was chosen to realize it through an additive manufacturing technology. Therefore, after testing a 3D printed polycarbonate prototype, the photo-camera support has been realized in Carbon Fiber (CF) Reinforced Composite material, exploiting one of the most innovative techniques emerging from the additive manufacturing global market. 
The CF structure was chosen in order to physically couple only materials with similar thermal and mechanical properties (the support plate of the chamber was made of CF as well). Moreover, it is sufficiently rigid to support the weight of the equipment (about 1 kg overall) without introducing deformations that may influence the  pointing on the target. 
An additional arm was added to it for a correct routing of the power and reading cables towards the bottom of the chamber to avoid interference with the beam. This support has been used for the 2019 MEG II engineering run. 

In order to improve stiffness of the support a new support has been realized in aluminium alloy 6060, always via additive manufacturing (Direct Metal Laser Melting process on a powder bed). This enhances also the related mounting screws pattern and reduces the effect of instantaneous deformation when the magnetic field is switched on.  The shape of the new aluminium support is the result of a topological optimisation  aimed at exploiting the larger rigidity of aluminium, but only introducing the strictly necessary material. This innovative interface has been mounted on the real setup in preparation for  the 2020 engineering run.

\begin{figure}[htbp]
\begin{center}
\includegraphics[width=0.45\textwidth]{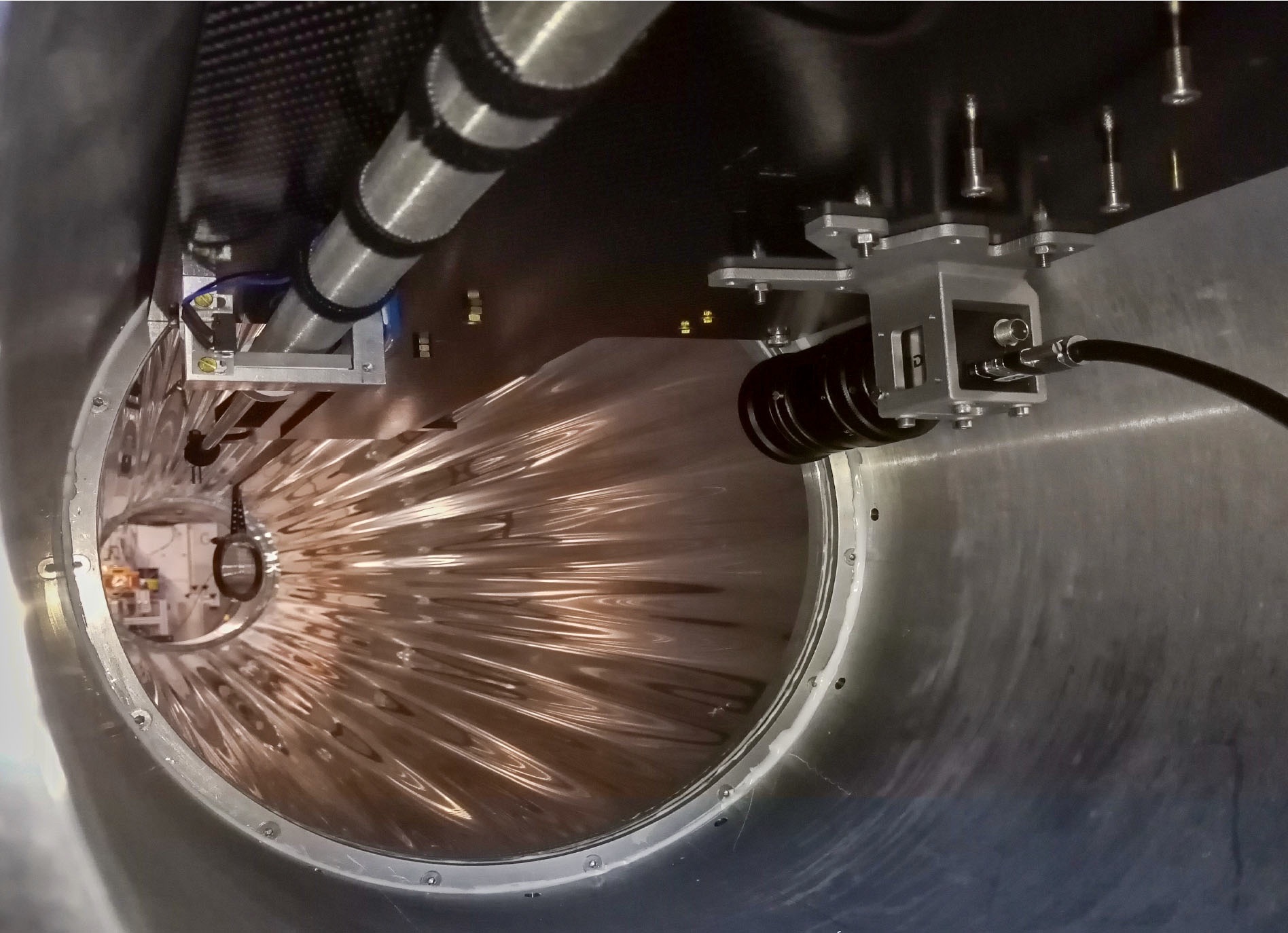}

\vspace{0.5cm}

\includegraphics[width=0.45\textwidth]{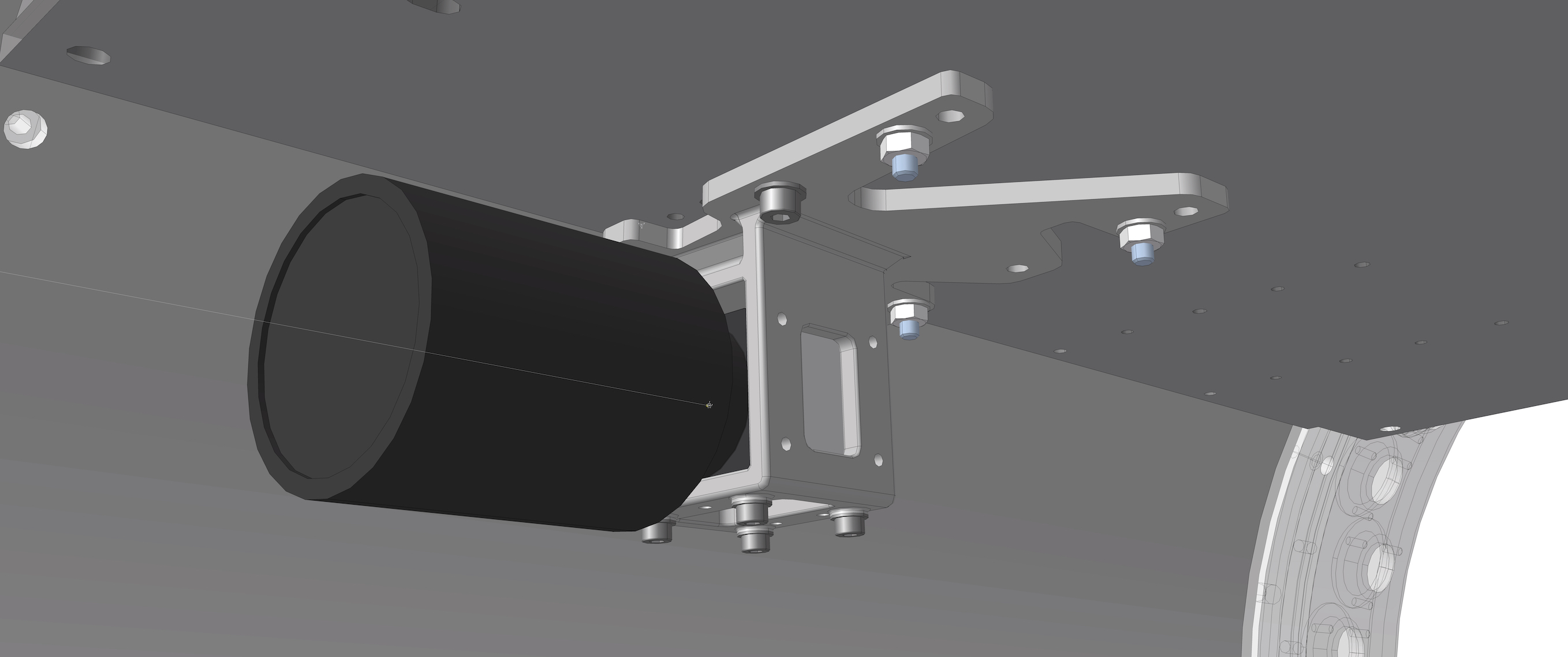}
\caption{Picture (upper plot) and CAD detail (lower plot) of the installation of the photo-camera with the final Al support in the inner cavity of the cylindrical drift chamber.}
\label{fig:installation}
\end{center}
\end{figure}

The support is fixed to the system for the target motion at a distance of about 1100~mm from the origin of the $z$ axis, in correspondence of the multi-wire chamber  end-plate. The transverse distance from the $z$ axis is about 120~mm,  with an angle of 6.3$^o$ with respect to the $z$ axis. 
As a result, the photo-camera frames an area of about 110~mm $\times$ 92~mm around the target center, which is  enough to image all the target and its support frame. 
 A picture of the photo-camera on the final Al support, installed in the MEG II detector and a CAD detail are shown in Fig.~\ref{fig:installation}. 

To have the largest possible portion of the target reasonably in focus, an aperture of f/16 was used, providing a large enough depth of field. Given these conditions, an exposure of 750~ms was chosen in order to optimize the use of the sensor's dynamic range for the best contrast. 

\subsection{\label{subsec:method}The method}
The  pattern  of dots can be reproduced by the photo-camera and the position of dots on the picture can be determined with standard image processing algorithms. If the target moves between two successive photo-camera shoots, 
the position of these patterns in the picture will change. A measurement of this displacement would allow to measure the corresponding
displacement of the target with respect to its original position. Given the size of the target to be imaged  and the resolution of our  photo-camera,
 one pixel in the image corresponds to a distance of a few tens $\mu$m on the target. Moreover, since imaging algorithms allow to reach a sub-pixel precision on the position of dot patterns, the goal 
of determining displacements below 100~$\mu$m in the transverse coordinates with respect to the optical axis 
is  within reach. Displacements along the optical axis can be detected considering that the distance $d$ between two
points on the target translates into a distance $d_I$ between two points on the image plane according to the magnification (M) formula:
\begin{equation}
\frac{d_I}{d} = M = \frac{f}{f - L}
\end{equation}
where $f$ is the focal length and $L$ is the distance of the target from the center of the photo-camera's optical system. Hence, 
a movement along the optical axis (i.e. a change of $L$) can be detected as a change of $d_I$. As we will show later, this 
approach can obtain the required resolution also for the coordinate transverse to the optical axis.

\section{The Target position measurement algorithm}

In this section we describe in detail the algorithms used to determine the dots positions within the photo-camera image and to use the measured
positions to extract the target position by means of a $\chi^2$ fit.

\subsection{Dot positions measurements}

\begin{figure}[htbp]
\begin{center}
\includegraphics[width=0.35\textwidth]{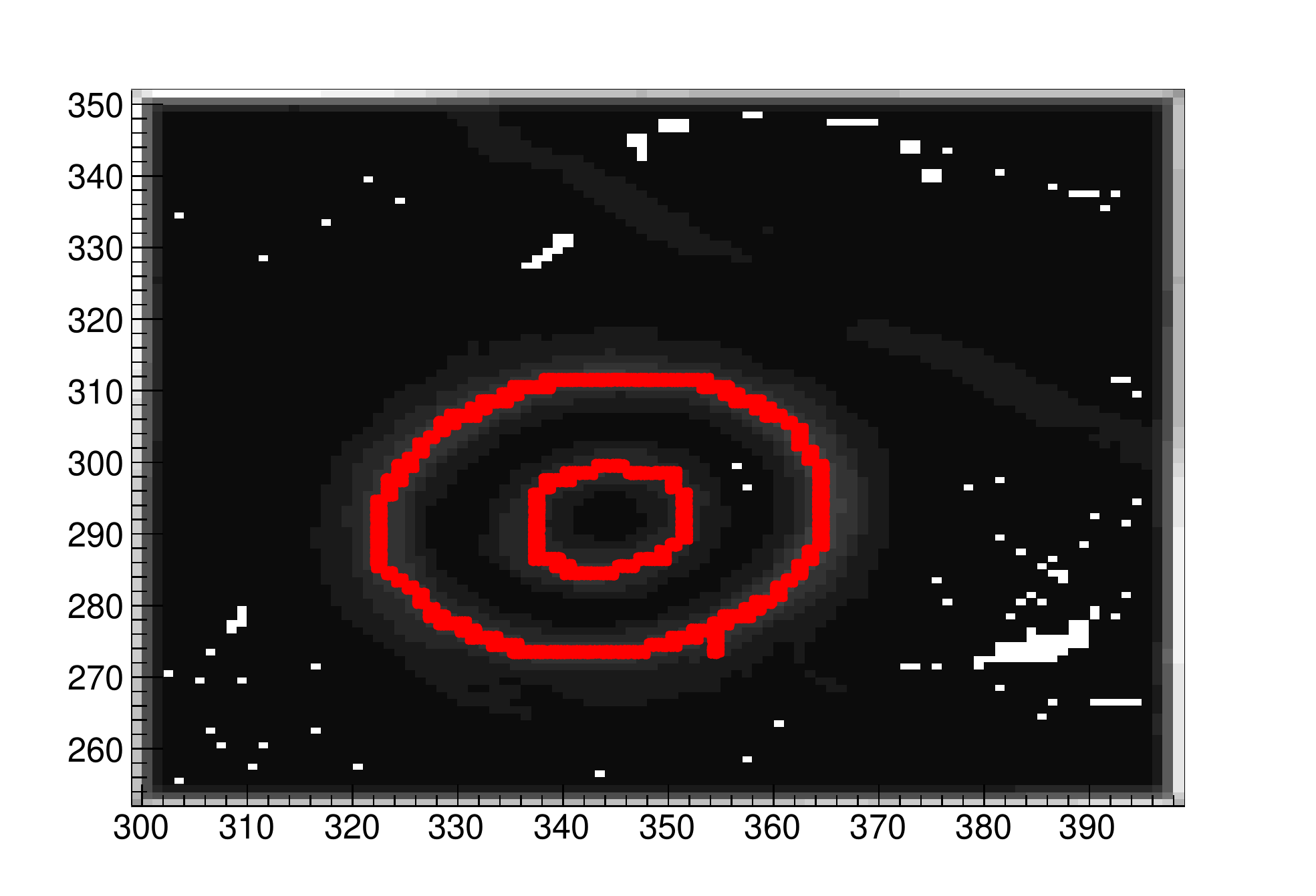}
\includegraphics[width=0.35\textwidth]{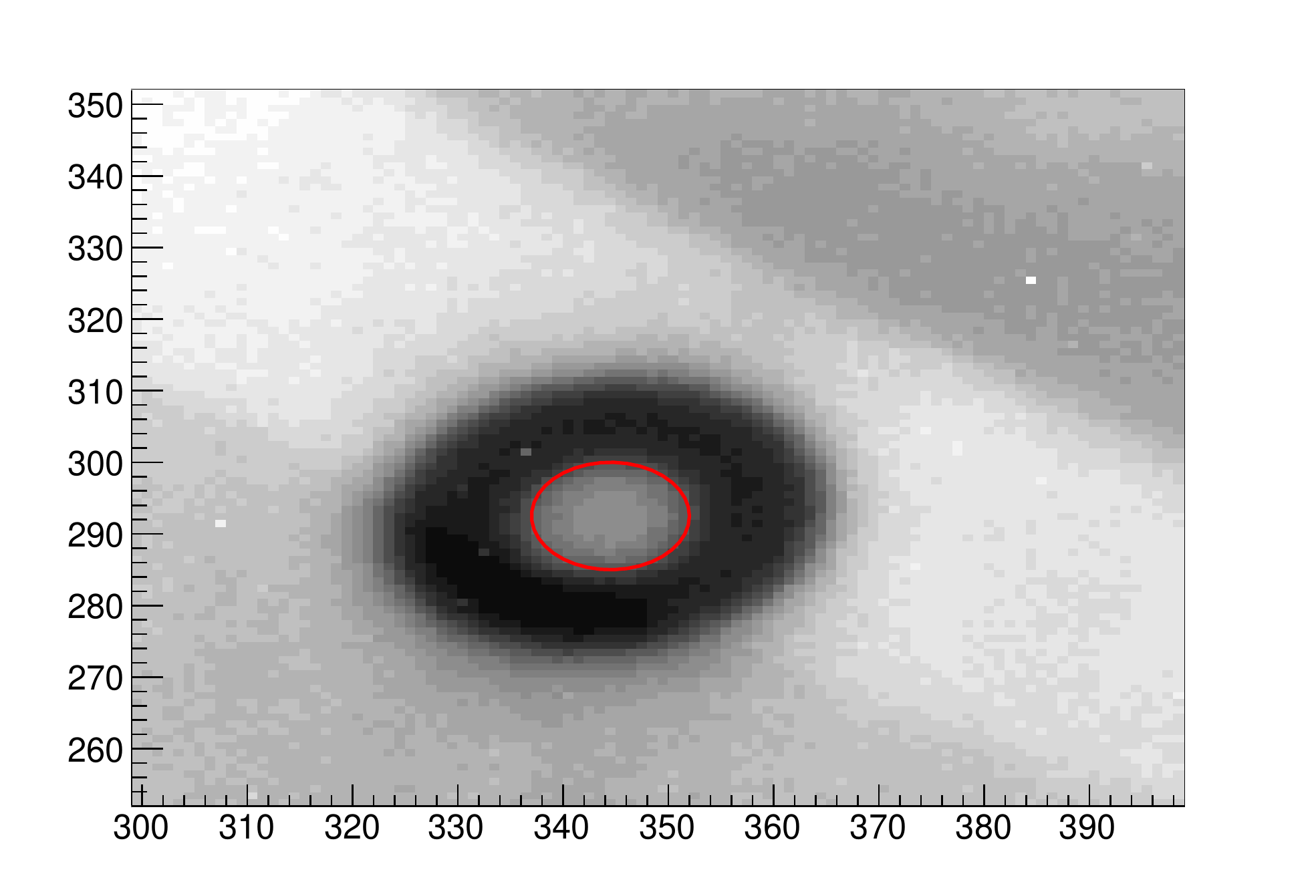}
\includegraphics[width=0.35\textwidth]{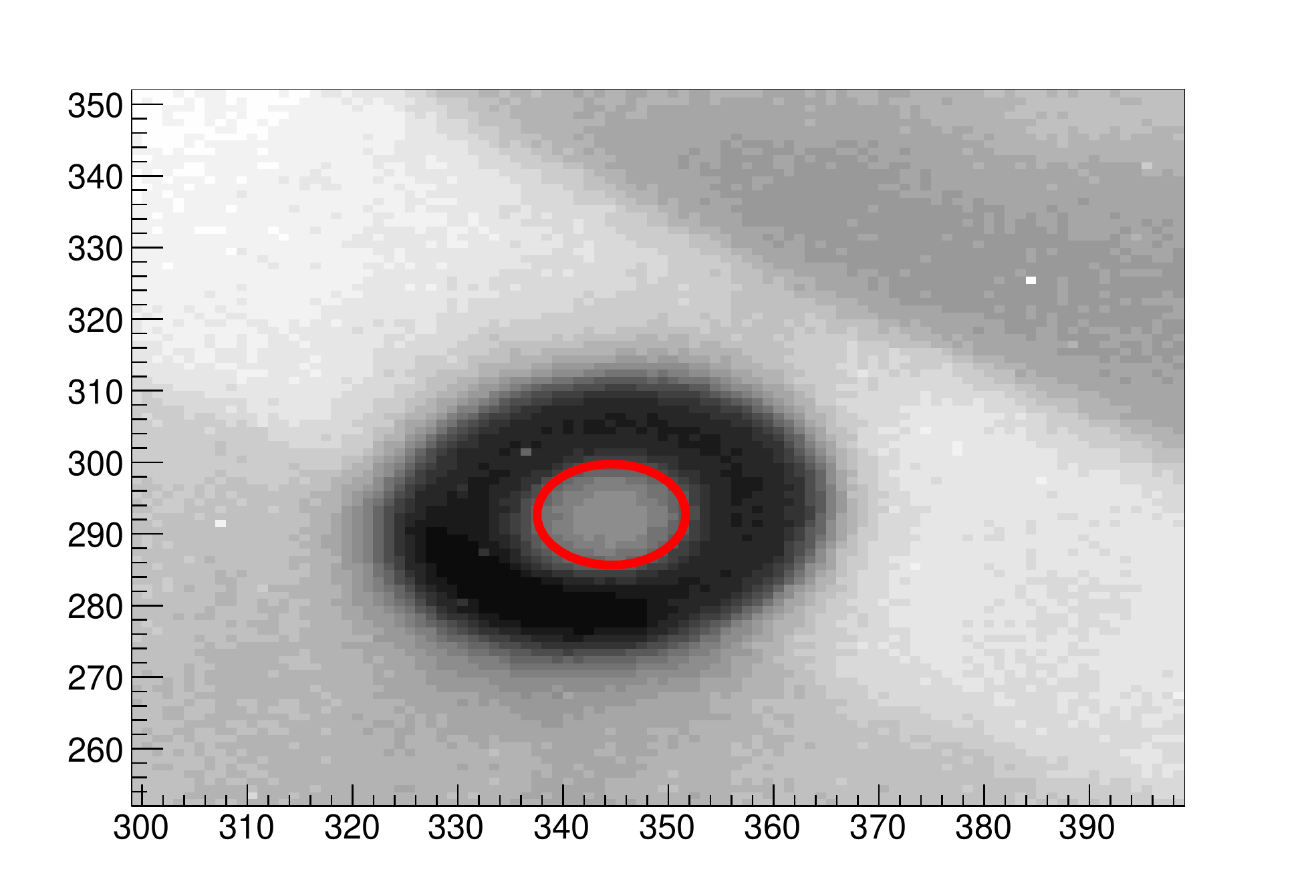}
\caption{Three-step procedure for the determination of the dot position: edge detection with the Canny algorithm (top); Identification of the white dot contour with a Hough transform (middle); $\chi^2$ fit for a precise determination of the dot center (bottom).}
\label{fig:dotfit}
\end{center}
\end{figure}

The dots positions are determined in a three-step procedure using standard image processing algorithms, as shown in Fig.~\ref{fig:dotfit}.
At first, a region of interest is automatically defined around each dot based on its expected position. A Canny edge detection algorithm~\cite{4767851} is applied to 
build an image of the dot edges. Secondly, a circular Hough Transform~\cite{Hough:1959qva} is applied to find which pixels belong to the edge between
the black contour and the white dot. Finally, a circumference is used to interpolate  the positions of these pixels with a $\chi^2$ minimization assuming 1 pixel uncertainy. The result of this fit procedure provides a 
measurement of the center of the white dot in the image. As an alternative approach, we evaluate a center of gravity of the picture light 
intensity to determine the center of the white dot, obtaining consistent results. 
\subsection{Target position and orientation measurement }

If the target and photo-camera positions are known, the positions of the dots on the photo-camera image can be inferred with simple arguments of geometrical optics.
In particular, given the center of the optical system, rays can be traced from a dot position on the target, through the optical center, to the sensor plane, giving the dot position in the image. With respect to the use of first-order optical relationships, this approach minimizes the systematic uncertainties. They might be   introduced by  the 
target inclination causing the dots far from the center of the
target to be  slightly out of focus.

The procedure can by formally described as an operator $\mathcal{T}$ acting on the 
3-dimensional position $\mathbf{r}_i = (x_i,y_i,z_i)$ of the real $i^{th}$ dot in the MEG II reference frame and producing a 2-dimensional 
position $\mathbf{s}_i = (p^x_i, p^y_i)$ on the sensor:
\begin{equation}
\mathbf{s}_i = \mathcal{T}(\mathbf{r}_i)
\end{equation}

The operator $\mathcal{T}$ has 7 parameters: the position of the optical center (3 parameters), the independent components of the unit vector
of the optical axis (2 parameters), the orientation of the sensor around the optical axis (1 parameter) and the distance of the sensor from the center of the optical system (1 parameter).
The positions $(p^x_i,p^y_i)$ are measured in units of number of pixels. 
The dots position in the MEG II reference frame, $\mathbf{r}_i$, can be derived from the dot positions in an arbitrarily defined target reference frame, 
$\mathbf{t}_i = (u_i,v_i,w_i)$:
\begin{equation}
\mathbf{r}_i = R \cdot \mathbf{t}_i + \mathbf{T}
\end{equation} 
where $R$ is a rotation matrix and $\mathbf{T}$ is a translation vector. If we place the center of the target reference frame at the center of 
the target, and we orient the first and second components of $\mathbf{t}_i$ along the major and minor axis, respectively, the vector
$\mathbf{T}$ gives the center of the target in the MEG II reference frame, while the matrix $R$ gives the target orientation. Hence,
the knowledge of the corresponding 6 parameters (the 3 components of the translation vector and the 3 Euler angles of the rotation matrix) is sufficient to measure the target position in the MEG II reference frame.

A $\chi^2$ function of the measured dot positions in the image with respect to the expected positions from the target orientation
and the geometrical optics can be defined as
\begin{equation}
\chi^2 = \sum_i \left[\mathbf{s}_i - \mathcal{T}(R \cdot \mathbf{t}_i + \mathbf{T})\right]^2
\end{equation}
The parameters of $\mathcal{T}$ and the dots position in the target reference frame can be inferred from
surveys performed at the beginning of the data taking period, as we will explain below, so that the parameters of $R$ and $\mathbf{T}$ (and hence the target position) can be determined by minimizing this $\chi^2$.

The target is assumed to be perfectly planar when installed ($w_i= 0$ for any $i$). If a deformation occurs during the MEG II data-taking run, 
it can be parameterized by an additional operator $\mathcal{Z}$ acting on the original positions $\mathbf{t}_i$. The $\chi^2$ becomes
\begin{equation}
\label{eq:chi2}
\chi^2 = \sum_i \left[\mathbf{s}_i - \mathcal{T}(R \cdot \mathcal{Z}(\mathbf{t}_i) + \mathbf{T})\right]^2
\end{equation}
and it will be minimized as a function of the parameters of $R$, $\mathbf{T}$ and $\mathcal{Z}$ in order to determine
the target position, its orientation and its deformation.

The operator $\mathcal{Z}$ is parameterized by means of the Zernike polynomials~\cite{1934MNRAS..94..377Z}, which are defined in a 2D system
of polar coordinates with $\rho \le 1$ as:
\begin{eqnarray}
Z_n^m(\rho,\phi) &=& R_n^m(\rho) \cos(m\phi) \\
Z_n^{-m}(\rho,\phi) &=& R_n^m(\rho) \sin(m\phi) 
\end{eqnarray}
where $m$ and $n$ are non-negative integers with $n \ge m$ and:
\begin{eqnarray}
R_{n}^{m}(\rho )=\sum _{k=0}^{\frac {n-m}{2}}{\frac {(-1)^{k}\,(n-k)!}{k!\left({\frac {n+m}{2}}-k\right)!\left({\frac {n-m}{2}}-k\right)!}}\;\rho ^{n-2\,k}
\end{eqnarray}
The first non-null radial polynomials are:
\begin{eqnarray}
R_0^0(\rho) &=& 1 \\
R_1^1(\rho) &=& \rho \\
R_2^0(\rho) &=& 2\rho^2 - 1 \\
R_2^2(\rho) &=& \rho^2 \\
R_3^1(\rho) &=& 3\rho^3 - 2\rho \\
R_3^3(\rho) &=& \rho^3
\end{eqnarray}

In the local $(u,v,w)$ reference frame, in order to describe a deformation of the target which is constrained to be null at the border 
thanks to the stiffness of the target frame, we define:
\begin{eqnarray}
\rho &=& \sqrt{(u/a)^2 + (v/b)^2} 
\end{eqnarray}
where $a$ and $b$ are the major and minor semi-axis of the target ellipse, and we use the following parameterization:
\begin{eqnarray}
\mathcal{Z}(u,v,w) = (u,v,w(u,v))
\end{eqnarray}
with:
\begin{eqnarray}
w(u,v) &=& \sum_{n,m} \left[ A_n^m \zeta_n^m(u,v) + A_n^{-m} \zeta_n^{-m}(u,v) \right] \\
\zeta_n^{\pm m}(u,v) &=& \frac{1}{2}\left[Z_{n}^{\pm m}(\rho,\phi) - Z_{n+2}^{\pm m}(\rho,\phi)\right]
\end{eqnarray}
The first term of the series is:
\begin{eqnarray}
w(u,v) = A_0^0 \cdot \frac{1}{2} \left[Z_0^0(\rho,\phi) - Z_2^0(\rho,\phi)\right] = A_0^0 \cdot (1 - \rho^2) \\
\end{eqnarray}
which describes a paraboloidal deformation.

\subsection{Operational procedure}

The positions of the dots in the target reference frame can be determined by a bench-top survey of the target foil, with an accuracy much better than 100~$\mu$m. Conversely, the position and orientation of the photo-camera (and hence the parameters of the operator $\mathcal{T}$) are not known with 
enough precision. To overcome this difficulty, 
we proceed as follows.
The target position at the beginning of a MEG II data-taking run will be precisely determined, with improved accuracy with respect to MEG, thanks to reflectors that are installed on the target frame for a laser survey. Immediately after, a set of pictures is taken (\emph{reference pictures}). We can assume that the position, orientation and shape of the target (and hence the parameters of $R$, $\mathbf{T}$ and 
$\mathcal{Z}$) are known for these pictures thanks to the recent surveys. So they can be fixed and the $\chi^2$ can be minimized with respect to the 7 parameters of $\mathcal{T}$. It provides a precise determination of these parameters.
When a new measurement of the target position is needed using the photogrammetric method, a new picture is taken and, in this case, the parameters of $\mathcal{T}$ are fixed from the reference fit, while the parameters of $R$, $\mathbf{T}$ and $\mathcal{Z}$ are fitted.

In order to make the procedure more robust against systematic effects associated to the inaccuracy of the optical model and to the initial
conditions of the target, when fitting the new pictures Eq.~(\ref{eq:chi2}) is in fact replaced by:
\begin{widetext}
\begin{equation}
\label{eq:chi2_rel}
\chi^2 = \sum_i \left\{(\mathbf{s}_i - \mathbf{s}^0_i) - \left[\mathcal{T}(R \cdot \mathcal{Z}(\mathbf{t}_i) + \mathbf{T}) - (\mathcal{T}(R^0 \cdot \mathcal{Z}^0(\mathbf{t}_i) + \mathbf{T}^0)\right]\right\}^2
\end{equation}
\end{widetext}
where $\mathbf{s}^0_i$ are measured and $R^0$, $\mathbf{T}^0$ and $\mathcal{Z}^0$ are fitted from the reference picture. In practice, the fit
to the target position is replaced by a fit to the target displacement. Anyway, the fitted parameters of $R$, $\mathbf{T}$ and $\mathcal{Z}$
are still referred to the global MEG II reference frame for an easy interface to the MEG II reconstruction software.

An estimate of the uncertainty in the measured position of the dots is given by the minimum $\chi^2$ divided by the number of degrees of freedom in the fit. It gives typically an error slightly below 1 pixel when the definition of Eq.~(\ref{eq:chi2}) is used. It includes any possible inaccuracy in the optical model and aplanarity of the target. The error goes down to $\frac{1}{3}$ of a pixel when Eq.~(\ref{eq:chi2_rel}) is used, demonstrating the superior robustness of this approach against these inaccuracies.

Typically a few dots per picture cannot be measured properly by the automatic procedure. They could be mostly recovered with a manual procedure by refining the regions of interest used to find and fit the dots. However,  their impact is so small that we decided to simply remove a dot from the fit when its contribution to the $\chi^2$ is larger than $25\sigma^2$, according to the estimate of the uncertainty described above.

It should be also stressed that several reference pictures need to be taken at the beginning of the data taking period, in order to reduce the statistical uncertainty on the estimate of the parameters of $\mathcal{T}$ to a negligible level. On the other hand, the $\chi^2$ defined in Eq.~\ref{eq:chi2_rel} requires a single reference picture. For this reason, we adopted the following procedure. One of the reference pictures is taken and used to get a preliminary estimate of the parameters of $\mathcal{T}$. This is used to fit for $R$, $\mathbf{T}$ and $\mathcal{Z}$ in the other reference pictures. Since the target did not move in between, one would expect to get zero displacements on average. Instead, statistical fluctuations in the first picture can be observed as an average fake displacement in the other pictures. The picture producing the minimum average displacements when used as the reference is taken as the single reference picture for the whole data taking period.

\section{Bench-top tests}

A test of the full procedure was performed by installing the photo-camera, a LED and a target mock-up on an optical table, with the target mounted on a  $2.5~\mu$m position accuracy linear stage.

\begin{figure}[htbp]
\begin{center}
\includegraphics[width=0.45\textwidth]{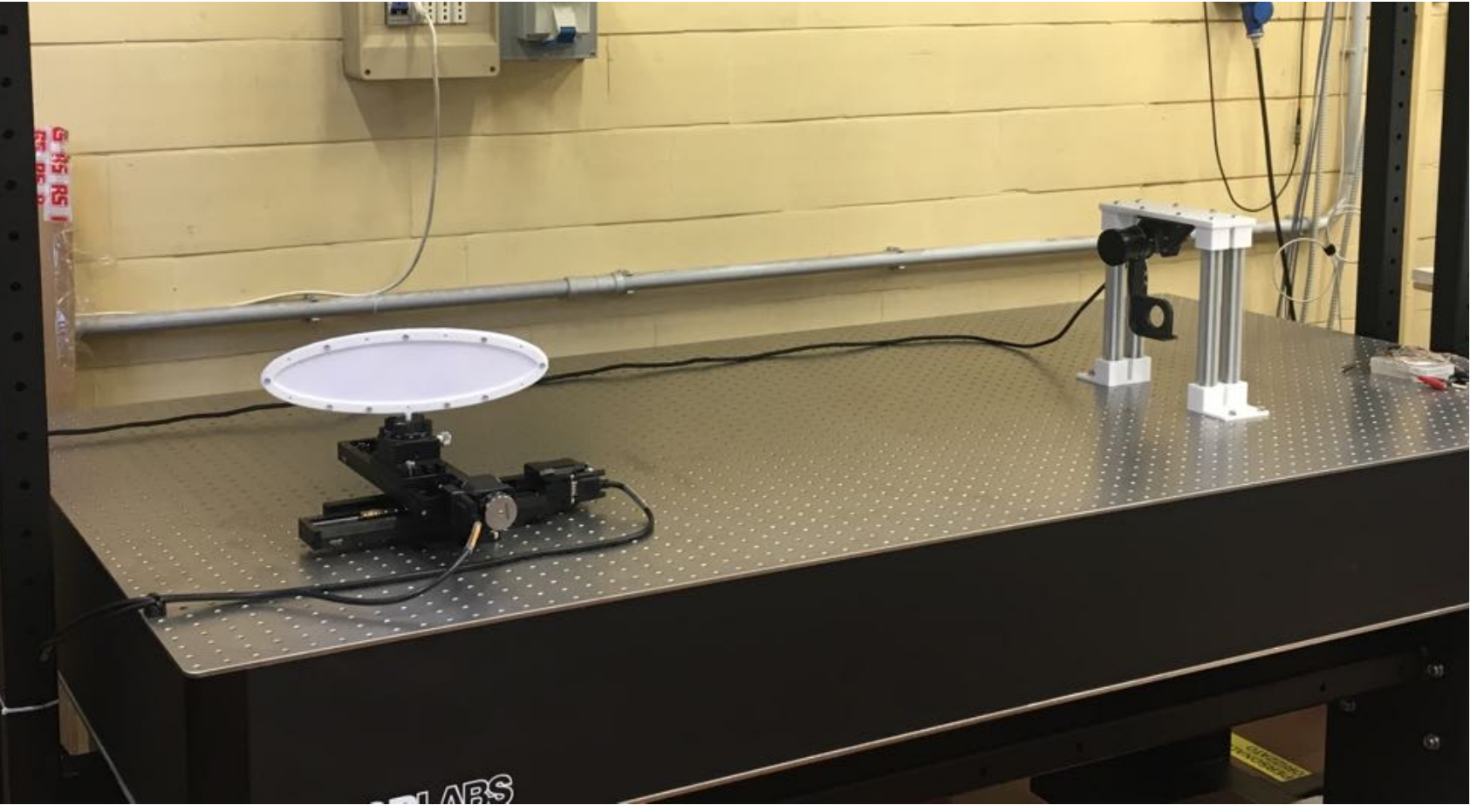}
\caption{Picture of the experimental setup for the bench-top test of the photogrammetric system.}
\label{fig:bench_setup}
\end{center}
\end{figure}

The assembly has been arranged in order to reproduce accurately the real setup installed inside the experiment. 
Exploiting 3D printing technologies available at INFN Roma Mechanical Workshop,  precise polycarbonate mechanical supports have been designed and 
produced. They are able to hold all the components in the correct relative positions between themselves 
and to interface properly  the optical table and the installed linear stage.  The photo-camera, instead, is fixed to the optical table using Al supports, in order to reduce thermal deformations.
A temperature sensor has been installed nearby the target for temperature monitoring while the environmental temperature is kept almost constant by air conditioning.
Fig~\ref{fig:bench_setup} shows the installed setup. 

A position scan  was performed independently along the $x$ and $z$ axes using the linear stages. This was used to  evaluate the precision to which shifts in the target position can be determined.

In this test  setup, we could not vary the target along the $y$ direction, but it should be noted that such movements have no impact on the track angles measurements in the MEG II experiment.

Before each scan, 10 pictures without moving the stages were taken, and one of them was chosen to serve as the reference picture, as described in the previous section.
In this setup, the initial coordinates of the target center are assumed to be (0,0,0), thus the fit returns  the coordinates $T_x$, $T_y$ and $T_z$ after the target movements. 

\begin{figure}[h]
\begin{center}
\includegraphics[width=0.45\textwidth]{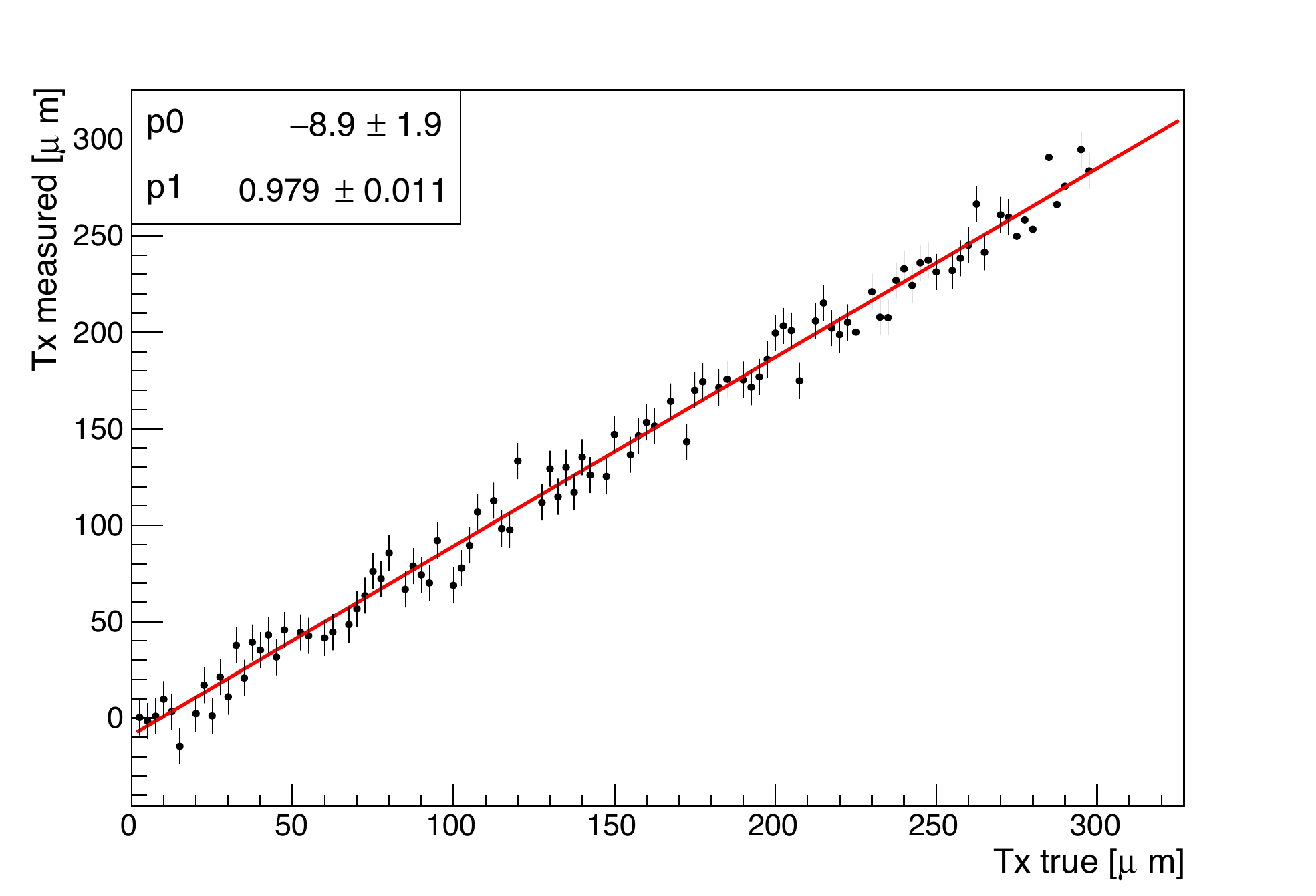}
\caption{$T_x$ fitted vs true $T_x$ in the $X$ scan described in the text. 
A linear fit is superimposed.}
\label{fig:scanxx}
\end{center}
\end{figure}
\begin{figure}[h]
\begin{center}
\includegraphics[width=0.45\textwidth]{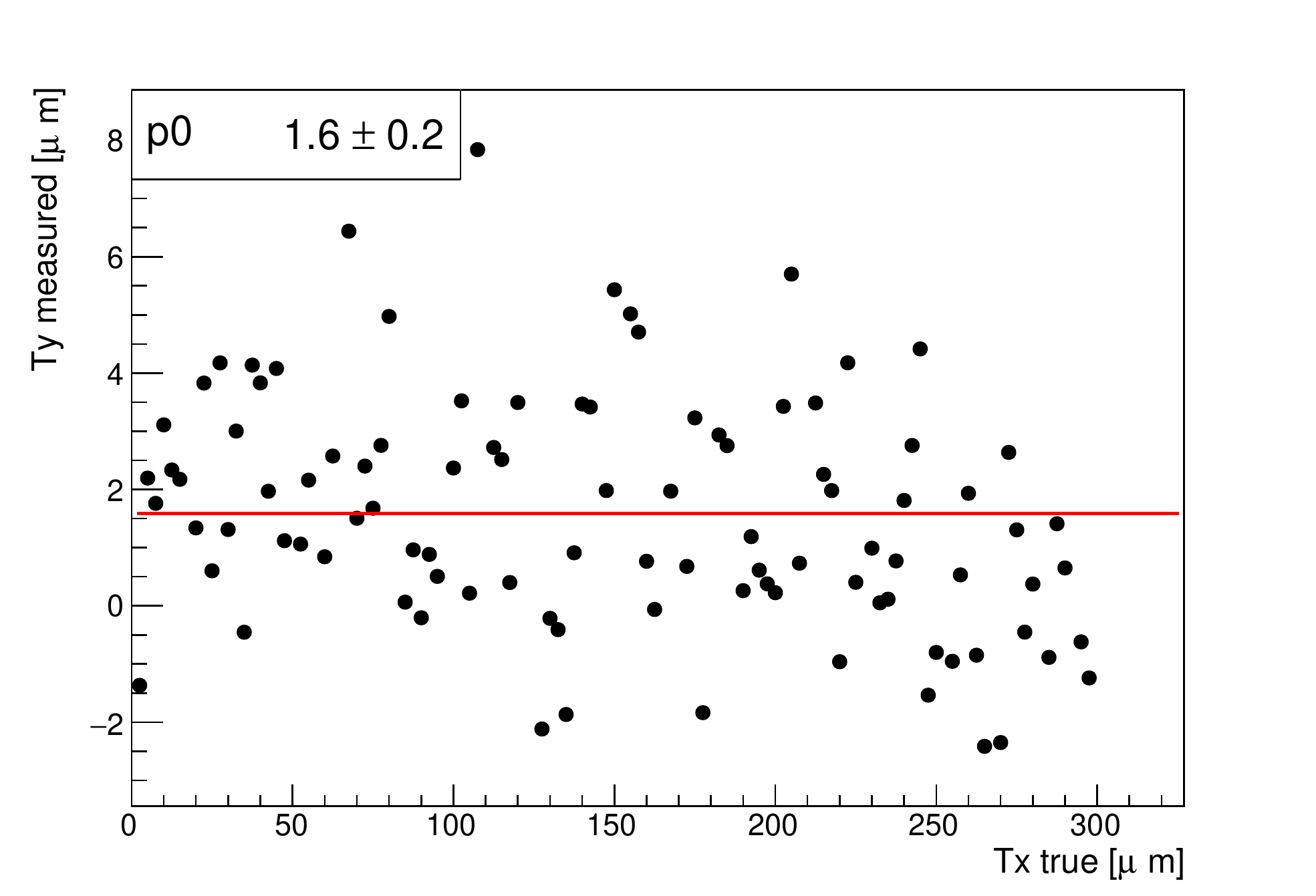}
\caption{$T_y$ fitted vs true $T_x$ in the $X$ scan described in the text.
A constant fit is superimposed.}
\label{fig:scanxy}
\end{center}
\end{figure}
\begin{figure}[h]
\begin{center}
\includegraphics[width=0.45\textwidth]{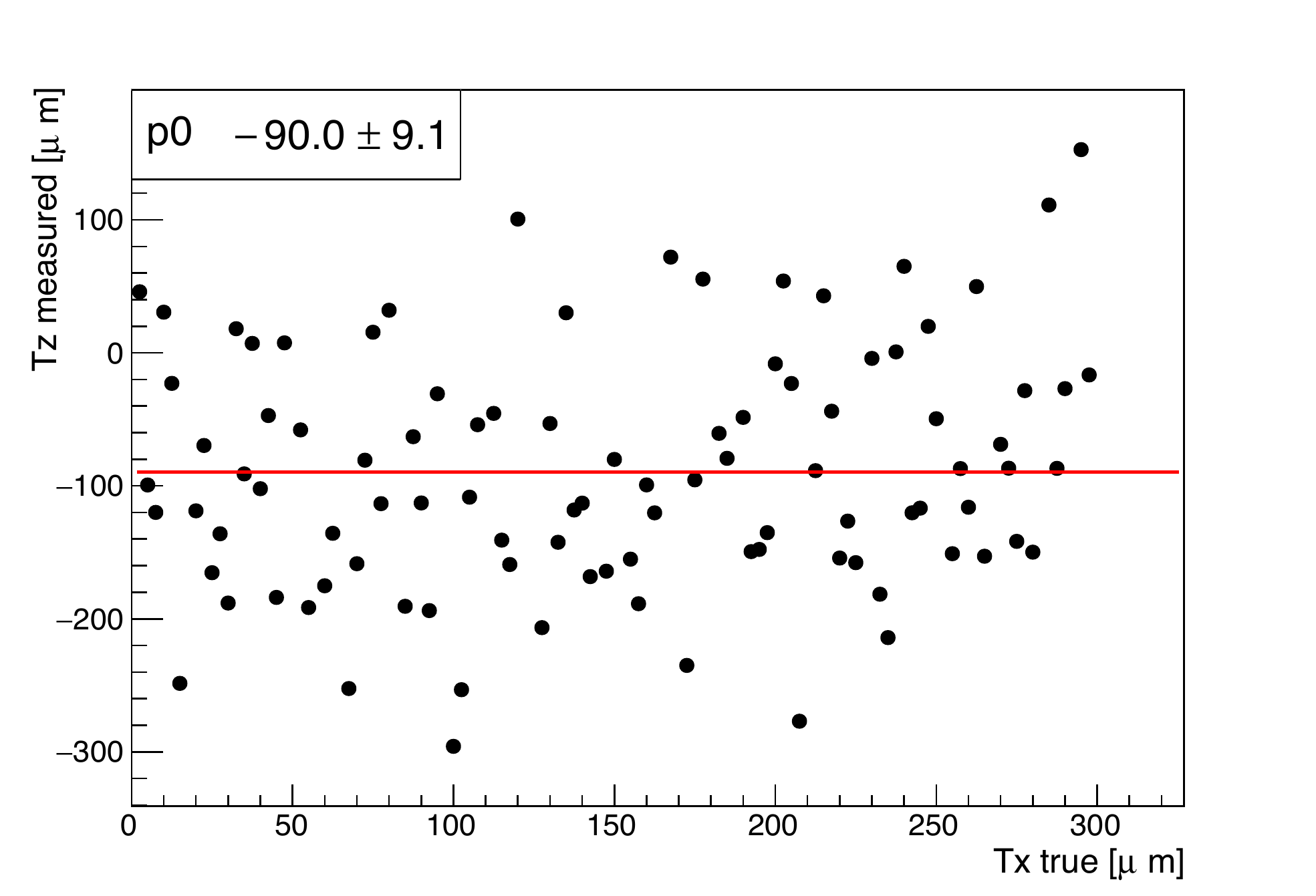}
\caption{$T_z$ fitted vs true $T_x$ in the $X$ scan described in the text. 
A constant fit is superimposed.}
\label{fig:scanxz}
\end{center}
\end{figure}

Figs ~\ref{fig:scanxx}, ~\ref{fig:scanxy}, ~\ref{fig:scanxz} show the fitted $T_x$, $T_y$ and $T_z$ as a function of the true $T_x$ in the $x$ scan. The pictures have been taken over $\sim6$ hours, in a random order with respect to the true shifts, so that time-dependent and shift-dependent biases mix incoherently and can be thus checked independently. Linear fits have been performed to the distributions and the errors on the fitted shifts have been estimated by mean of a linear regression in the case of the fitted $T_x$. 
The resulting uncertainty on $T_x$ is $\sigma(T_x) = 12~\mu$m. Given that the direction transverse to the target plane is almost coincident with the $X$ axis we can conclude that we fully satisfy our precision requirements.
The angular coefficients and the intercept are consistent with one and zero, as expected.
A bias in $T_z$ is observed, that is significantly different from 0 but still within the requirements. It is probably due to the residual uncertainty of the reference picture.

\begin{figure}[htbp]
\begin{center}
\includegraphics[width=0.45\textwidth]{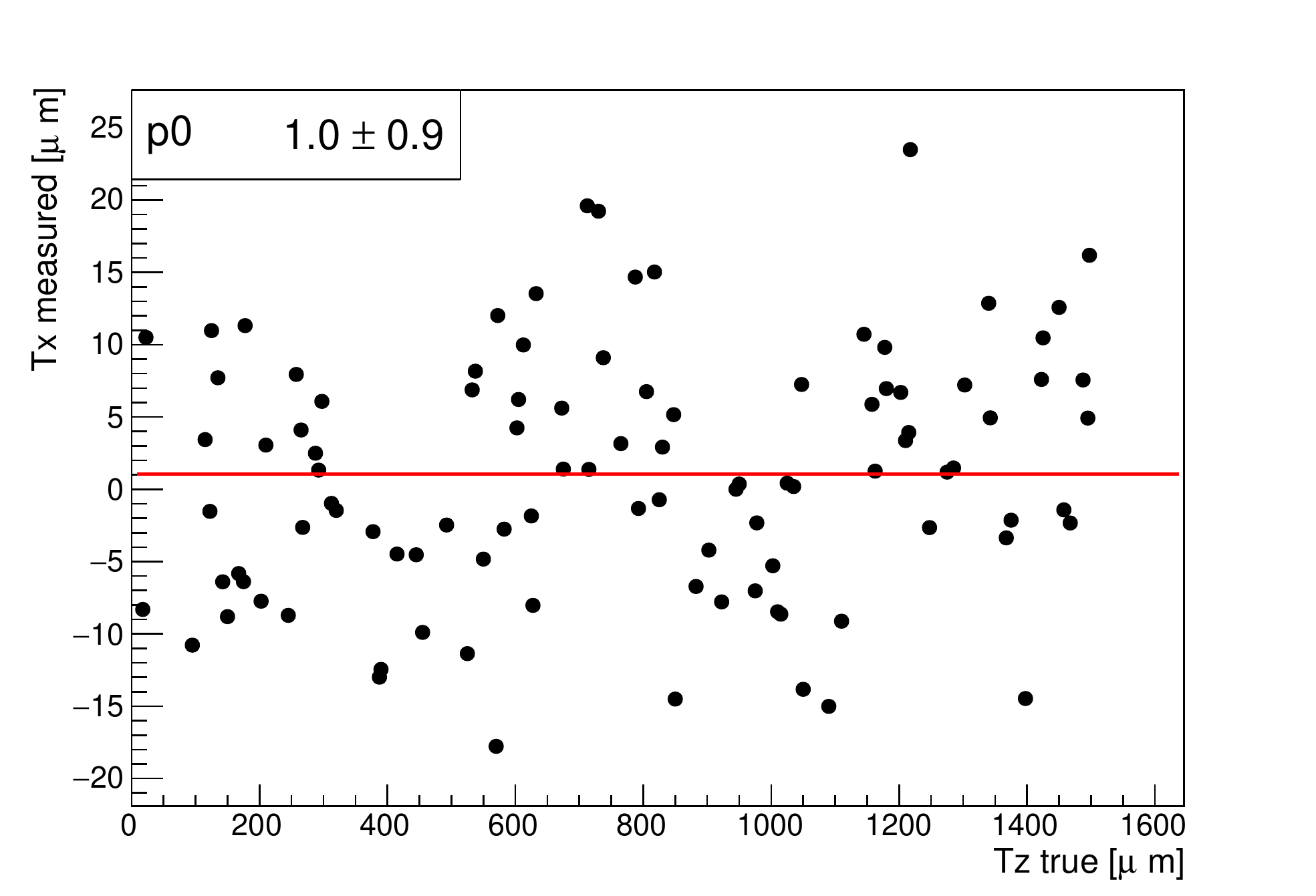}
\caption{$T_x$ fitted vs true $T_z$ in the $Z$ scan described in the text. 
A constant fit is superimposed.}
\label{fig:scanzx}
\end{center}
\end{figure}

\begin{figure}[htbp]
\begin{center}
\includegraphics[width=0.45\textwidth]{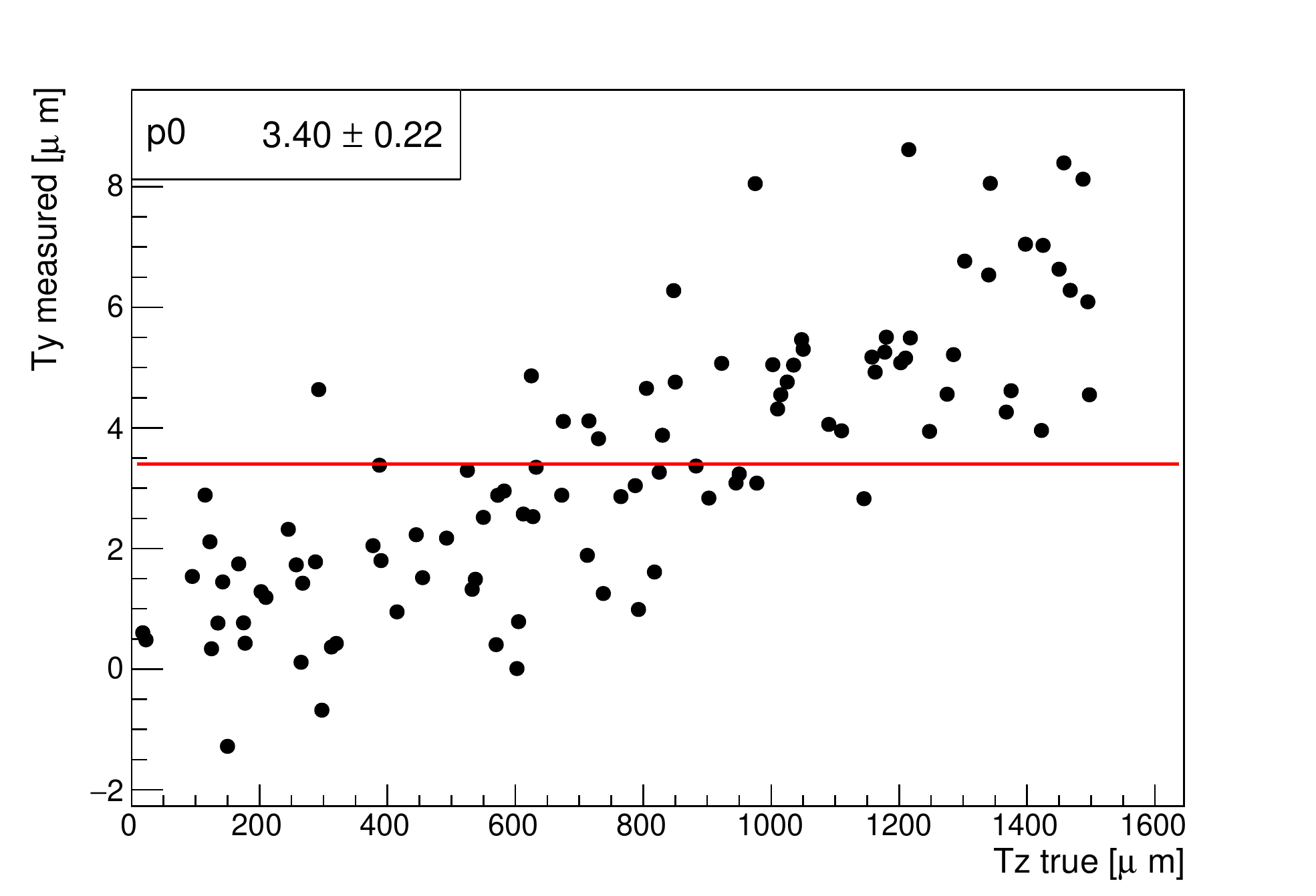}
\caption{$T_y$ fitted vs true $T_z$ in the $Z$ scan described in the text. 
A constant fit is superimposed.}
\label{fig:scanzy}
\end{center}
\end{figure}

\begin{figure}[htbp]
\begin{center}
\includegraphics[width=0.45\textwidth]{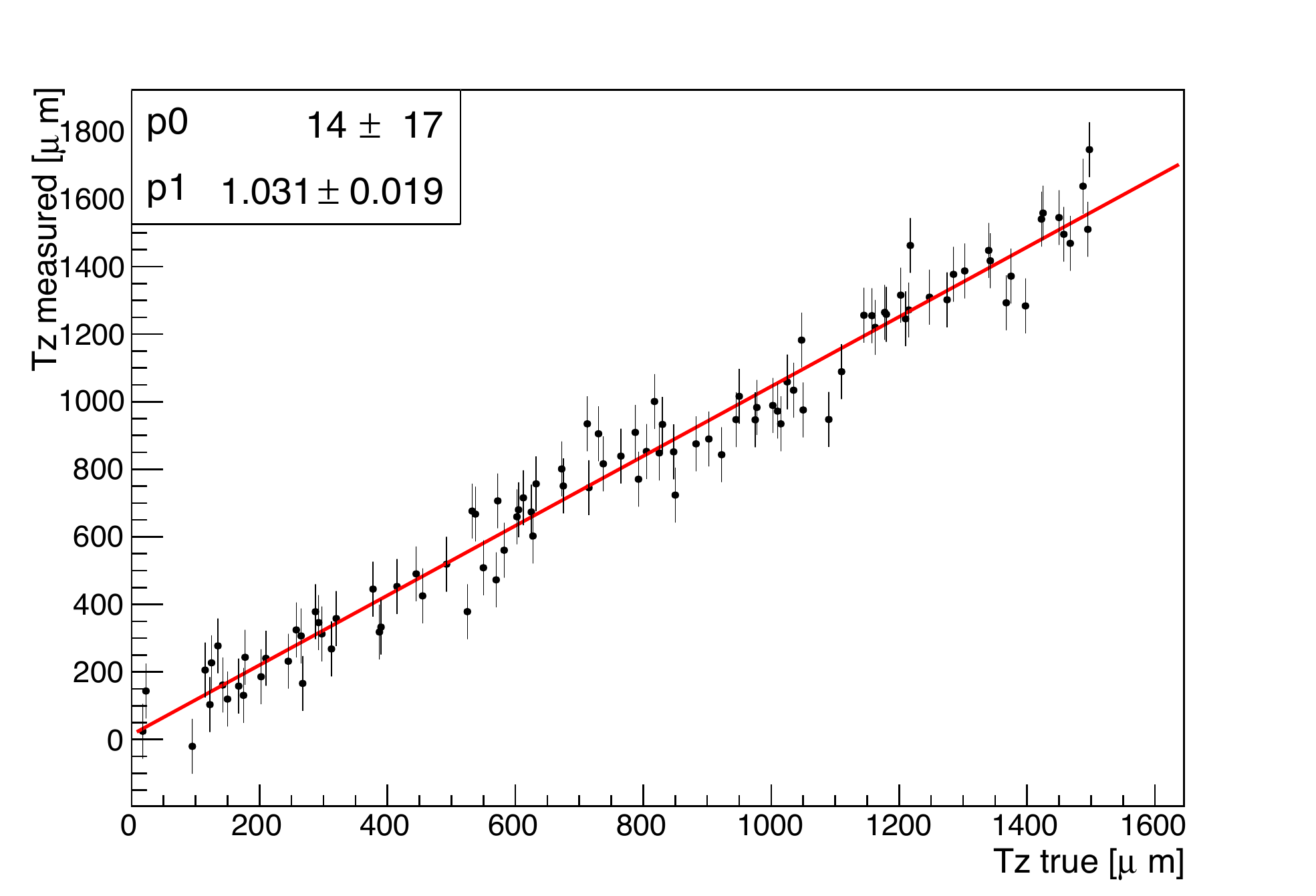}
\caption{$T_z$ fitted vs true $T_z$ in the $Z$ scan described in the text. 
A linear fit is superimposed.}
\label{fig:scanzz}
\end{center}
\end{figure}

Figs ~\ref{fig:scanzx}, ~\ref{fig:scanzy}, ~\ref{fig:scanzz} show the fitted $T_x$, $T_y$, $T_z$ as a function of the true $T_z$ in the $Z$ scan. The pictures have been taken over $\sim6$ hours, again in a random order with respect to the true shifts.
Linear fits have been performed to the distributions and the error on the fitted $T_z $ has been estimated by mean of a linear regression.
The resulting uncertainty on $T_z$ is  $\sigma(T_z) =82~\mu$m. The angular coefficients and the intercept are consistent with one and zero, as expected, also in this case. 

\begin{figure}[htbp]
\begin{center}
\includegraphics[width=0.45\textwidth]{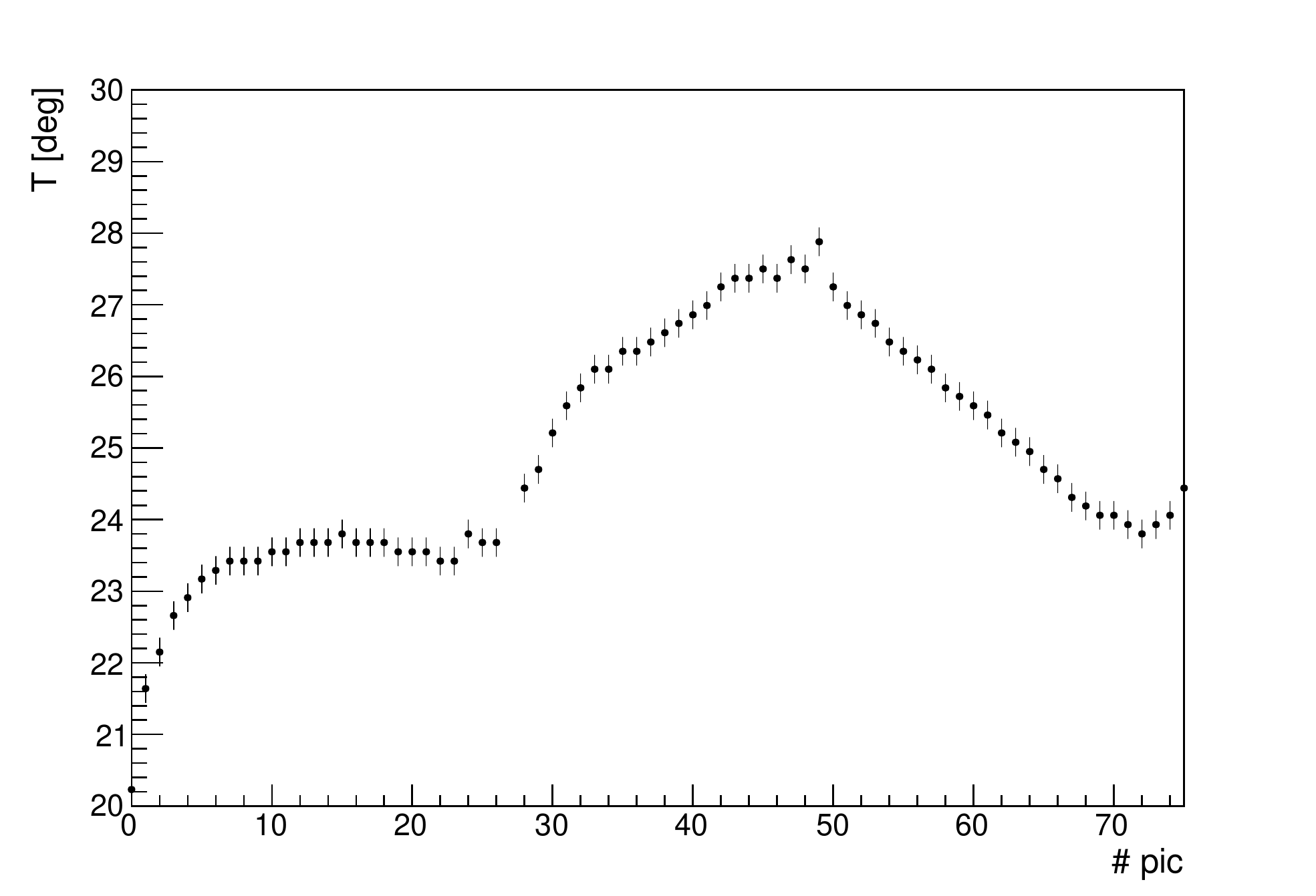}
\caption{Temperature vs time during the data-taking without moving the stages.}
\label{fig:temp}
\end{center}
\end{figure}

\begin{figure}[htbp]
\begin{center}
\includegraphics[width=0.45\textwidth]{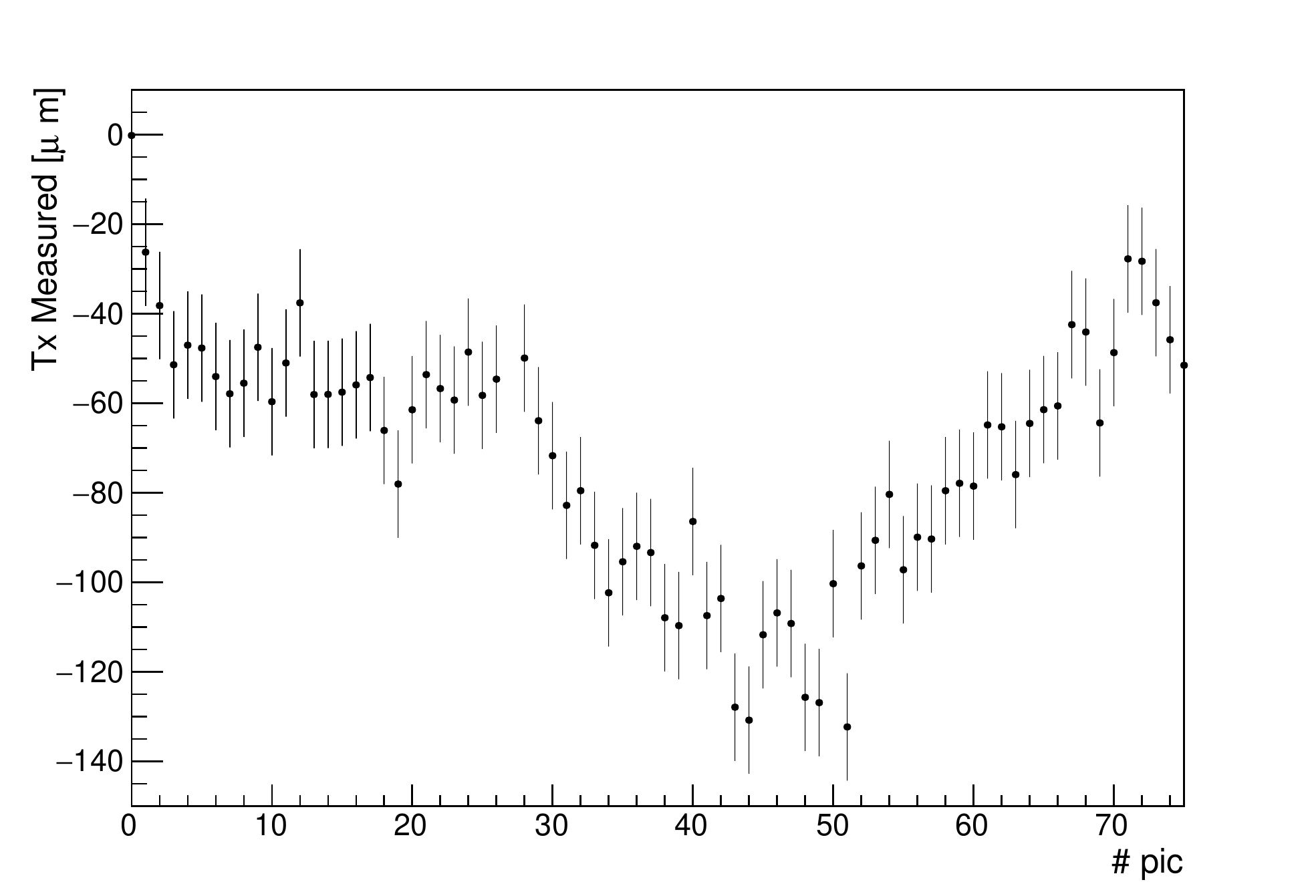}
\includegraphics[width=0.45\textwidth]{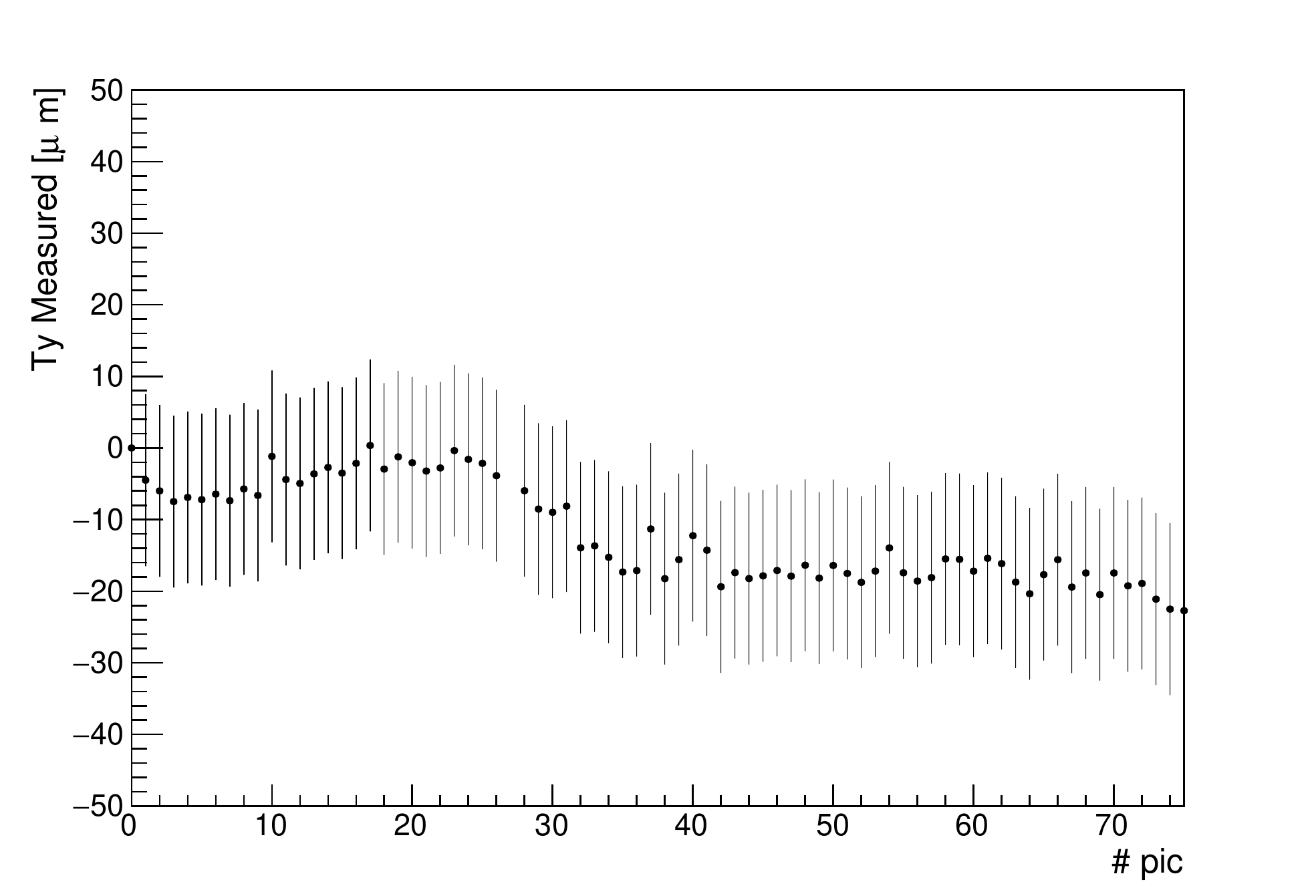}
\includegraphics[width=0.45\textwidth]{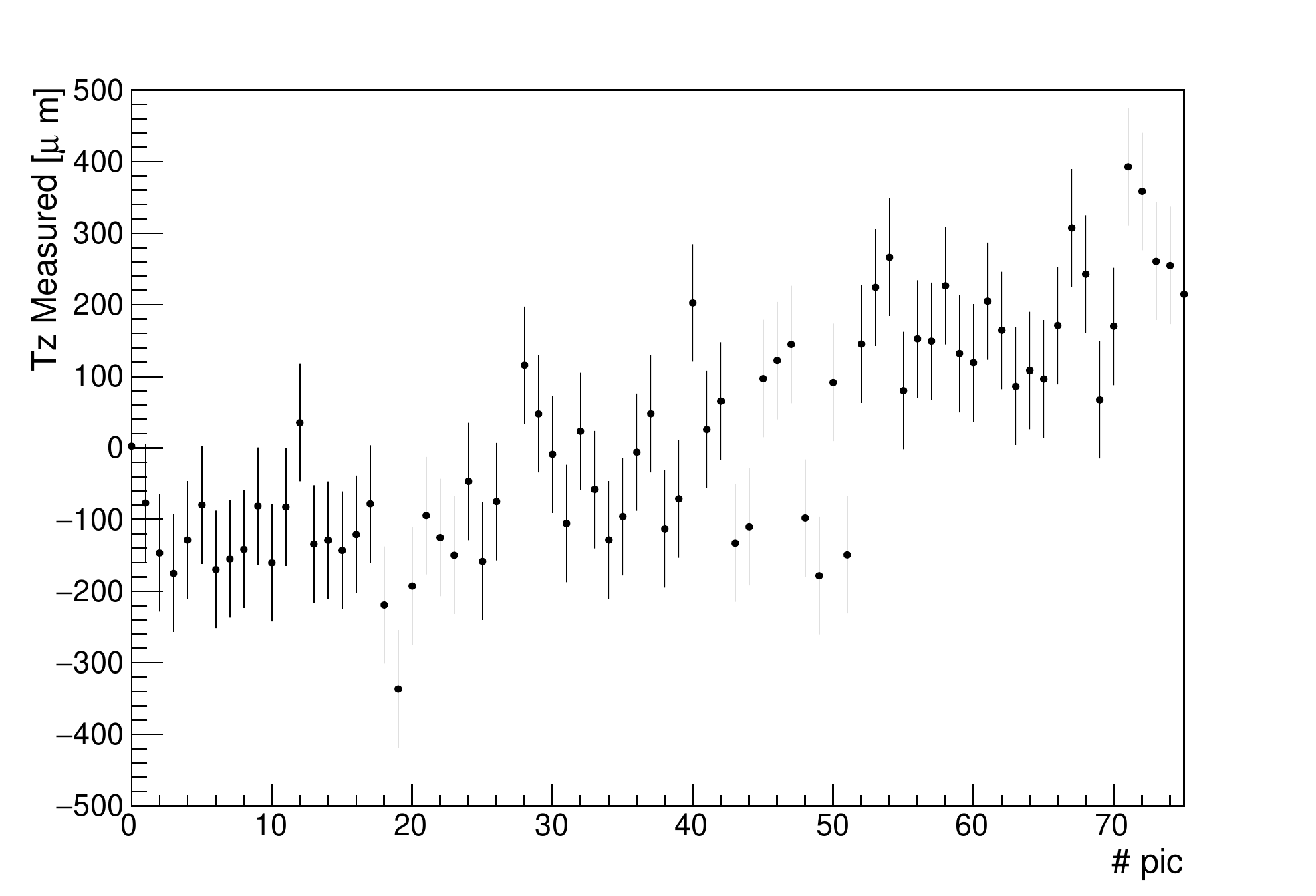}
\caption{Fitted $T_x$ (upper plot) $T_y$ (middle plot) $T_z$ (lower plot) for the pictures taken without moving the stages.}
\label{fig:NoSlitXYZ}
\end{center}
\end{figure}

The dependencies of the fitted position of the target as a function of the environmental temperature changes has been observed  by taking 75 pictures in 30 minutes without moving the stages.   Figure~\ref{fig:temp} shows the variation of the temperature versus time, during the data taking period while Figs ~\ref{fig:NoSlitXYZ} show the fitted $T_x$, $T_y$, $T_z$ from the reference picture. In these figures, the errors estimated from the $x$ and $z$ scans described previously have been assumed on $T_x$ and $T_z$ while the error on $T_y$ has been assumed equal to the error in $T_x$ although this is probably an overestimation, as it was computed in a configuration where the stage was moved.

We observed a correlation of the shifts to the temperature, that is clear in $T_x$ but can be also perceived in $T_y$ and $T_z$. Looking at fixed reference points in the background of the pictures, we concluded that this dependence is due to real deformations of the
target, not a change in the photo-camera position or response. Unfortunately, the poor stiffness of the target frame used in this test prevents to fit the deformations with the Zernike polynomials approach described above. Moreover, we cannot know a priori the amount of deformation induced by the temperature changes. It makes not possible to give a robust estimate of the sensitivity to these deformations. Nonetheless, this result clearly demonstrates the possibility of monitoring this kind of effects with a precision below 100~$\mu$m in all coordinates.

A study of the correlations among the fitted parameters was also performed. An example of the correlation matrix extracted from one of the fit is shown in Tab.~\ref{tab:corr}. Very large correlations are observed among some parameters, owing to the misalignment between the optical axis and the $z$ axis. Indeed, we checked that correlations among single parameters are small if the fit is performed 
in a reference frame aligned with the optical axis, and emerge when the parameters are combined to get the position in the MEG II reference frame. These effects need to be taken into account when calculating the resolution for displacements along the normal direction to the target plane. An uncertainty propagation that includes the correlation between $T_x$ and $T_z$ gives for instance a resolution of 32~$\mu$m for displacements along the direction normal to the target plane, the most dangerous for the positron track angle measurements. A calculation of the eigenvalues and eigenvectors of the covariance matrix does not give indications of directions in the parameter space along which there is very poor resolution (\emph{weak modes}).

\begin{table*}
    \caption{\label{tab:corr} Example of correlation matrix for a displacement of the target of $55~\mu$m along the $X$ axis with respect to the reference position. Translations are described by the three vector components of $\mathbf{T}$. Rotations are described by three Euler angles according to the conventions used in the MEG II software.}
    \centering
    \begin{ruledtabular}
    \begin{tabular}{ccccccc}
    & $T_x$ & $T_y$ & $T_z$ & $\theta_1$ & $\theta_2$ & $\theta_3$ \\ 
    \hline
       $T_x$ & 1.000 & -0.022 & 0.983 & 0.012 & 0.787 & -0.006 \\
       $T_y$ & -0.022 & 1.000 & -0.022 &  0.005 & -0.015 & -0.010 \\
       $T_z$ & 0.983 & -0.022 & 1.000 & 0.015 & 0.799 & -0.011 \\
       $\theta_1$ & 0.012& 0.005 & 0.015 & 1.000 & -0.001 & -0.778 \\
       $\theta_2$ &  0.787 & -0.015 & 0.799& -0.001 & 1.000 & 0.006 \\
       $\theta_3$  & -0.006 & -0.010 & -0.011 & -0.778 & 0.006 & 1.000 \\
    \end{tabular}
    \end{ruledtabular}
    \label{tab:my_label}
\end{table*}
 
\section{Measurements in the MEG II experiment}
\begin{figure}[htbp]
\begin{center}
\includegraphics[width=0.45\textwidth]{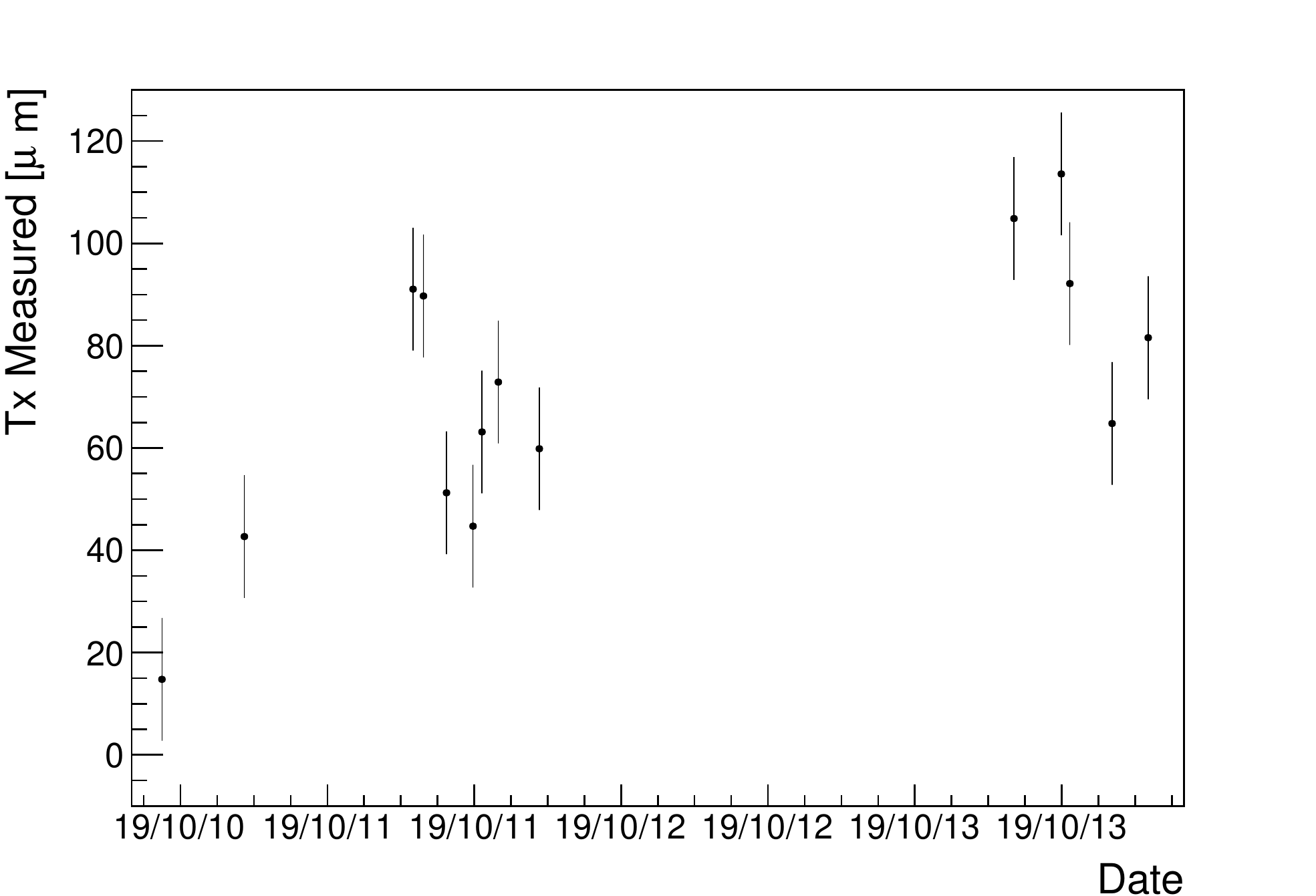}
\includegraphics[width=0.45\textwidth]{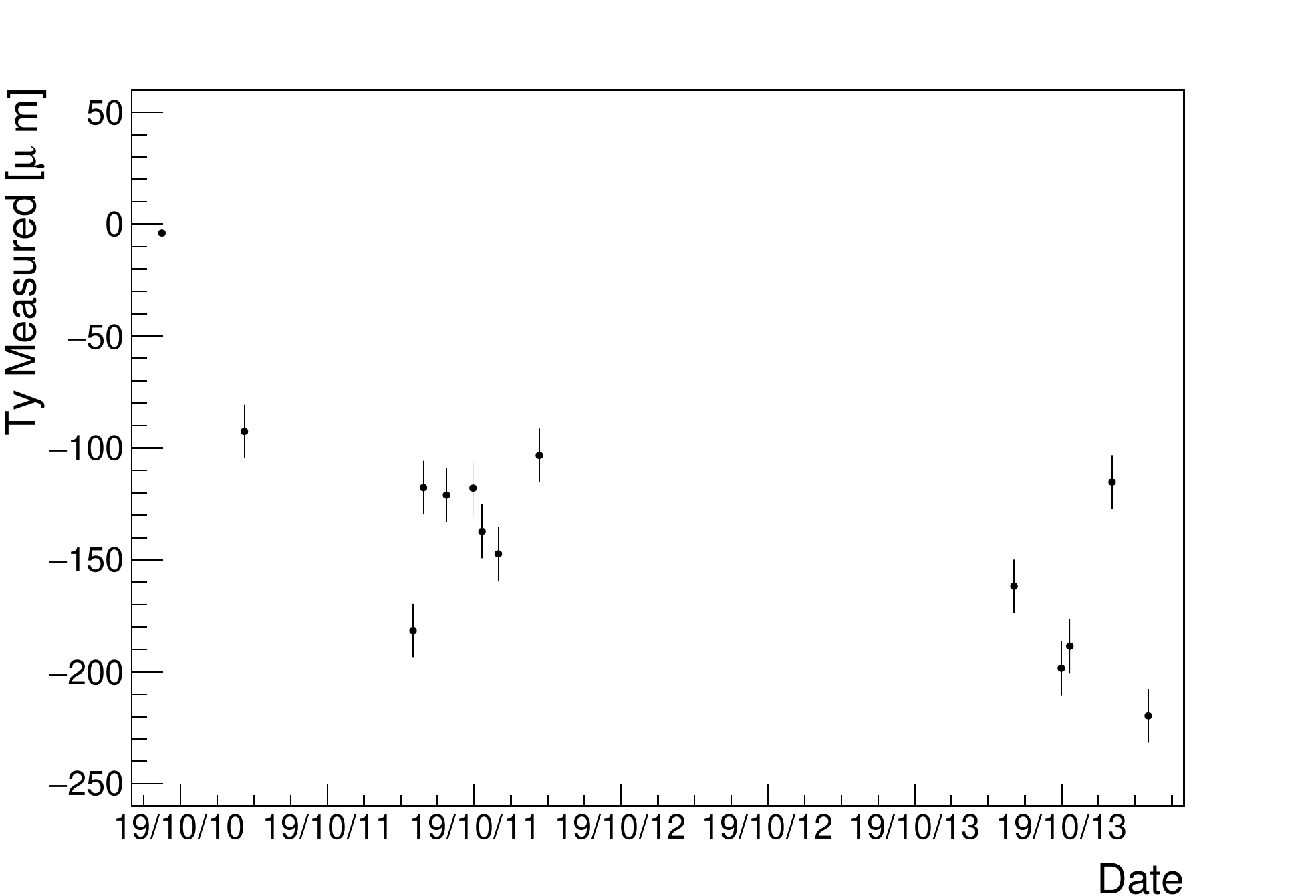}
\includegraphics[width=0.45\textwidth]{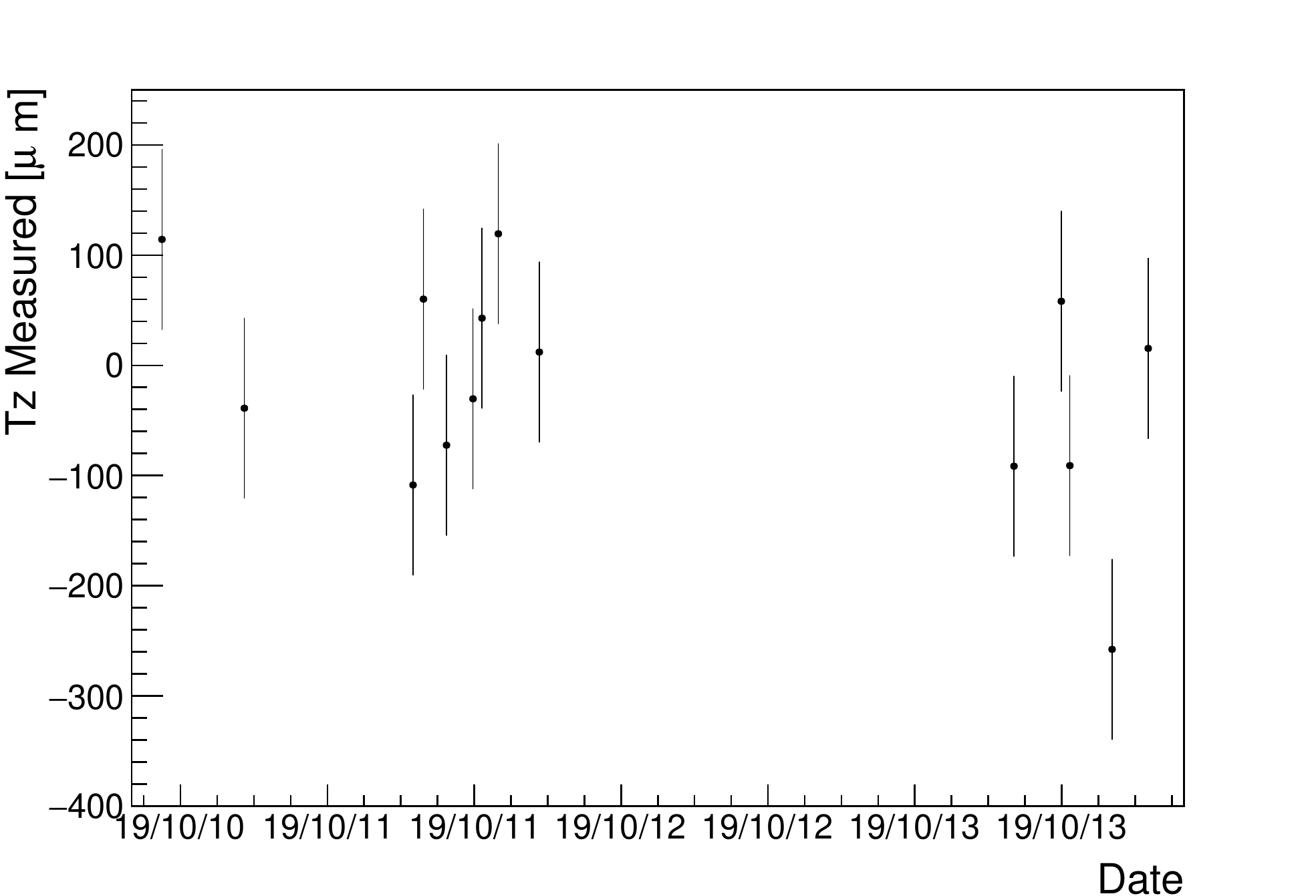}
\caption{Fitted $T_x$,$T_y$ and $T_z$ for pictures taken during  2019 MEG II engineering run.}
\label{fig:run2019}
\end{center}
\end{figure}

We operated successfully the photo-camera during the MEG II 2018 and 2019 engineering runs.
As an example, fit results for $T_x$, $T_y$, $T_z$, assuming the reference position in (0,0,0), are shown in Fig.~\ref{fig:run2019} for a time interval of one day. The time interval with no measurement corresponds to cycles of extraction and insertion of the target.
\section{Conclusions}
A photogrammetric method for the monitoring of the target position during the MEG II data-taking  is presented. The method exploits imaging
techniques to find displacements of patterns drawn on the target with respect to a reference picture taken at the beginning of a data-taking run. By combining this
information with the results of an optical survey , it is possible to determine the position of
the target during the run, when the target is not accessible. The method described reaches the required resolution of less than 100~$\mu$m on the displacements along the axis normal to  the target plane.

The photo-camera system has to be permanently installed inside the MEG II magnetic field volume and operated with the magnetic field on. Hence, it has been designed  to avoid
the presence of any ferromagnetic component. Moreover, a USB communication interface has been selected to avoid failure observed with an Ethernet interface during the first engineering data-taking run. Finally, the photo-camera will be placed at a sufficient distance from the beam axis in order  not to  interfere with the beam 
halo. All these features have been tested during the engineering MEG II  runs in 2017, 2018 and 2019.

A bench-top test has been performed at INFN Roma with the same photo-camera of the final system, in a geometrical
arrangement which reproduces the set-up inside the MEG II magnetic field. The accuracy of the measurement of the target displacement with respect to a reference picture
has been measured to be $\sigma(\Delta x) = $12 $\mu$m, and  $\sigma(\Delta z) = $82 $\mu$m.
Even in the worse situation of a large displacement of a few mm along the 
optical axis, the accuracy remains below the MEG II requirements. We also notice that the performances 
are significantly affected by the presence of large correlations between displacements along $x$ and $z$. This could be significantly improved 
by combining the images of two photo-cameras looking at the target from two different points of view. 

All these results make highly recommendable the installation of the system in the final setup of the MEG II experiment, with no evident interference with the rest of the apparatus. Eventually, a two-photo-camera analysis will be developed to improve the performances.\\

\section*{Acknowledgments}
We are grateful to our colleagues from the MEG II collaboration for their support in the development of this study.

\section*{Data Availability}
The data that support the findings of this study are available from the corresponding author upon reasonable request.

\bibliography{aipsamp}

\nocite{*}

\end{document}